\documentclass[11pt]{article}
\usepackage{epsfig,color,fullpage,citesort}
\def\myendproof{{\ \vbox{\hrule\hbox{%
   \vrule height1.3ex\hskip0.8ex\vrule}\hrule }}\par}
\newtheorem{theorem}{Theorem}[section]
\newtheorem{lemma}[theorem]{Lemma}

\newenvironment{proof}{{\it Proof. }}{\myendproof}
\newcommand{\qed}{\myendproof}
\newcommand{\setof}[1]{\{{#1}\}}
\newcommand{\set}[2]{\{{{#1}:{#2}}\}} 
\newcommand{\fname}[1]{{\sc #1}}
\newcommand{\note}[1]{}
\newcommand{\maxof}[1]{\max\setof{#1}}
\newcommand{\notchb}{\not\chb}
\newcommand{\chb}{\rightarrow}
\newcommand{\opt}[1]{h({#1})}

\newcommand{\cost}[1]{{c({#1})}}

\newcommand{\height}[1]{{h({#1})}}

\newcommand{\fall}[1]{{\tilde{h}(#1)}}

\newcommand{\comment}[1]{}
\newcommand{\Xomit}[1]{}
\newcommand{\first}[2]{\mbox{\it first}_{#1}(#2)}
\newcommand{\init}{\bot}
\newcommand{\fini}{\top}
\newcommand{\pr}[2]{\mbox{\it pred}_{#1}(#2)}
\newcommand{\su}[2]{\mbox{\it succ}_{#1}(#2)}
\newcommand{\prr}[1]{\mbox{\it pred}(#1)}
\newcommand{\suu}[1]{\mbox{\it succ}(#1)}

\newcommand{\head}[1]{\mbox{\it start}(#1)}
\newcommand{\tail}[1]{\mbox{\it end}(#1)}
\newcommand{\algo}[1]{{\sc{#1}}}
\newcommand{\cut}{\Gamma}
\newcommand{\jopt}[1]{h_j(G[{#1}])}
\newcommand{\decomp}[1]{\mbox{\algo{Decomp}}(#1)}
\newcommand{\merge}[1]{\mbox{\algo{Merge}}(#1)}
\newcommand{\minheight}[1]{\mbox{\algo{MinHeight}}(#1)}
\newcommand{\best}[1]{\mbox{\algo{Best}}(#1)}
\newcommand{\cuttrans}[1]{\mbox{\algo{CutTrans}}(#1)}
\newcommand{\chainpair}[1]{\mbox{\algo{ChainPair}}(#1)}
\newcommand{\construct}[1]{\mbox{\algo{Construct}}(#1)}
\newcommand{\removerange}[1]{\mbox{\algo{RemoveRange}}(#1)}
\newcommand{\newcut}[1]{\mbox{\algo{Newcut}}(#1)}

\title{Detecting Race Conditions in Parallel Programs that Use
Semaphores\thanks{Preliminaries versions of this paper appeared
in~\cite{Lu:19xx:DRC,Klein:19xx:RCD}.%
}}
\author{Philip N. Klein\thanks{Department of Computer Science, Brown
     University, 
     Providence, RI 02912, USA. Email:
     klein@cs.brown.edu.}
\and
  Hsueh-I Lu\thanks{Corresponding author. Institute of Information Science, Academia
  Sinica, Taipei 115, Taiwan. Email: hil@iis.sinica.edu.tw. URL: www.iis.sinica.edu.tw/\~{ }hil/ }
\and
  Robert H.B. Netzer\thanks{Department of Computer Science, Brown University, 
     Providence, RI 02912, USA. Email:
     rn@cs.brown.edu.}
}

\begin{document}
\maketitle
\begin{abstract}
We address the problem of detecting race conditions in programs that
use semaphores for synchronization. Netzer and Miller showed that it
is NP-complete to detect race conditions in programs that use many
semaphores. We show in this paper that it remains NP-complete even if
only two semaphores are used in the parallel programs.

For the tractable case, i.e., using only one semaphore, we give two
algorithms for detecting race conditions from the trace of executing a
parallel program on $p$ processors, where $n$ semaphore operations are
executed.  The first algorithm determines in $O(n)$ time whether a
race condition exists between any two given operations. The second
algorithm runs in $O(np\log n)$ time and outputs a compact
representation from which one can determine in $O(1)$ time whether a
race condition exists between any two given operations. The second
algorithm is near-optimal in that the running time is only $O(\log n)$
times the time required simply to write down the output.
%
%This paper combines the results in the two preliminary
%versions~\cite{Lu:19xx:DRC,Klein:19xx:RCD}.
%
%{\em Keywords:} Parallel debugging, Synchronization, Race conditions,
%Semaphores, NP-completeness, Polynomial-time algorithms, Scheduling
\end{abstract}

\section{Introduction}
Race detection is crucial in developing and debugging shared-memory
parallel
programs~\cite{Simmons:1996:DPT,Savage:1997:EDD,Emrath:1992:DNP,Netzer:1992:WRC,Itzkovitz:1999:TID,Ha:2002:SEF}. Explicit
synchronization is usually added to such programs to coordinate access
to shared data.  For example, when using a semaphore, a $V$-operation
increments the semaphore, and a $P$-operation waits until the
semaphore is greater than zero and then decrements the
semaphore. $P$-operations are typically used to wait (synchronize)
until some condition is true (such as a shared buffer becoming
non-empty), and $V$-operations typically signal that some condition is
now true.  Race conditions result when this synchronization does not
force concurrent processes to access data in the expected order.  One
way to dynamically detect races in a program is to trace its execution
and analyze the traces afterward.  A central part of dynamic race
detection is to compute from the trace the order in which
shared-memory accesses were guaranteed by the execution's
synchronization to have executed. Accesses to the same location not
guaranteed to execute in some particular order are considered a race.
When programs use semaphore operations for synchronization, some
operations (belonging to different processes) could have potentially
executed in an order different than what was traced.

In this paper, we address the tractability of detecting race
conditions from the traces of parallel programs that use semaphores.
Let $p$ be the number of processors used to execute the parallel
program, and let $n$ be the total number of semaphore operations
performed in the execution. The trace can then be represented by a
directed $n$-node graph $G$ consisting of $p$ disjoint chains, each
represents the sequence of semaphore operations executed by a
processor.  A {\em schedule} of $G$ is a linear ordering of all nodes
in $G$ consistent with the precedence constraints imposed by the arcs
of $G$.  A prefix of a schedule of $G$ is a {\em subschedule} of $G$.
A subschedule of $G$ is {\em valid} if at each point in the
subschedule, the number of $V$ operations is never exceeded by the
number of $P$ operations for each semaphore (i.e., all semaphores are
always nonnegative).  Then, if the trace indicates that $v$ preceded
$w$ in the actual execution, but a valid subschedule\footnote{We
consider subschedules rather than schedules because deadlocks might
happen during the execution of parallel programs.}  exists in which
$w$ precedes $v$, then $v$ and $w$ could have executed in either
order, i.e., there is a {\em race condition} between $v$ and $w$.
Miller and Netzer showed that detecting race conditions in parallel
programs that use multiple semaphores is
NP-complete~\cite{Netzer:1990:CEO}.  Researchers have developed exact
algorithms for cases where the problem is efficiently solvable
(programs that use types of synchronization weaker than semaphores
such as
post/wait/clear)~\cite{Netzer:1992:ERC,Helmbold:1993:CSO,Helmbold:1996:TRC},
and heuristics for the multiple semaphore
case~\cite{Emrath:19xx:ESA,Helmbold:19xx:ATA}.  The complexity for the
case of constant number of semaphores was unknown. In the present
paper, we show that the problem remains NP-complete even if only two
semaphores are used in the parallel program.

For the case of using only one semaphore in parallel programs, we give
two algorithms.  The first algorithm detects in $O(n)$ time whether a
race condition exists between any two operations.  The second
algorithm computes in $O(np\log n)$ time a compact representation,
from which one can determine whether a race condition exists between
any two operations in $O(1)$ time.  Our results are based on the
reducing the problem of determining whether a valid subschedule exists
in which $w$ precedes $v$ to the problem of {\em Sequencing to
Minimize Maximum Cumulative Cost (SMMCC)\/}.
%We first describe the SMMCC problem and then explain the
%equivalence in the next two paragraphs.  
Given an acyclic directed graph $G$ with costs on the nodes, the {\em
cumulative cost} of the first $i$ nodes in a schedule of $G$ is the
sum of the cost of these nodes.  Thus, minimizing the maximum
cumulative cost is an attempt to ensure that the cumulative cost stays
low throughout the schedule. The SMMCC problem is NP-complete in
general even if the node costs are restricted to
$\pm1$~\cite{Abdel-Wahab:1976:SAR,Garey:1979:CIG}.  Abdel-Wahab and
Kameda~\cite{Abdel-Wahab:1978:SMM} presented an $O(n^2)$-time
algorithm for the special case that $G$ is a series-parallel graph.
(The time bound was later improved to $O(n \log n)$ by the same
authors~\cite{Abdel-Wahab:1980:SOS}.)  As part of this solution, they
gave an $O(n\log p)$-time algorithm applicable when $G$ consists of
$p$ disjoint chains.  The existence problem of a valid {\em schedule}
in which $v$ precedes $w$ can be reduced to the SMMCC problem in a
chain graph augmented with one inter-chain edge. We add an edge from
$w$ to $v$, assign costs to the nodes ($+1$ if the node is a
$P$-operation, $-1$ if a $V$-operation), and compute the minimum
maximum cumulative cost.  Clearly, the cost is non-positive if and
only if there is a valid schedule. The augmented chain graph is not
series-parallel, so the algorithms of Abdel-Wahab and
Kameda~\cite{Abdel-Wahab:1978:SMM,Abdel-Wahab:1980:SOS} are not
applicable.  We show that the SMMCC problem can nevertheless be solved
in polynomial time.  In fact, for the special case of interest, that
in which the costs are $\pm 1$, we give a linear-time algorithm.

The rest of the paper is organized as follows.
Section~\ref{sec:prelim} gives the preliminaries.
Section~\ref{sec:single} gives the algorithm for a single pair of
nodes. Section~\ref{sec:all} gives the algorithm for all pairs of
nodes. Section~\ref{sec:2semaphores} sketches the proof for showing that
race-condition detection is NP-complete if two semaphores are used in
the parallel program.

\section{Preliminaries}
\label{sec:prelim}
%\subsection{Definition and Notation}
Suppose $G$ is an acyclic graph with node costs.  We introduce some
terminology having to do with schedules, mostly adapted
from~\cite{Abdel-Wahab:1978:SMM}.  
%A {\em schedule} of $G$ is a
%sequence of $G$'s nodes which is consistent with the precedence
%constraints imposed by the arcs of $G$. 
A {\em segment} of a schedule is a consecutive subsequence.  Let $H =
v_1v_2\cdots v_m$ be a sequence of nodes.  The {\em cost} of $H$,
denoted $\cost{H}$, is the sum of the costs of its nodes.  The {\em
height of a node $v_\ell$ in $H$} is defined to be the sum of the
costs of the nodes $v_1$ through $v_\ell$.  The {\em height of $H$},
denoted $\height{H}$, is the maximum of 0 and the maximum height of
the nodes in $H$.
%(a) the maximum height of any node in $H$, if
%some node of $H$ has non-negative height, or (b) zero, if all nodes in
%$H$ have negative heights.  
A node of maximum height in $H$ is called a {\em peak}. A node of
minimum height in $H$ is called a {\em valley}.  The {\em reverse
height} of $H$, denote $\fall{H}$, is the height of $H$ minus the cost
of $H$.  Note that height and reverse height are nonnegative.  A
schedule of $G$ is {\em optimal} if its height is minimum over all
schedules of $G$.  We use $\opt{G}$ to denote the height of its
optimal schedule.

A sequence $C=v_1v_2\cdots v_m$ of nodes of $G$ is called a {\em chain}
of $G$ if the only edges in $G$ incident on these nodes are $v_0v_1,
v_1v_2,\ldots,v_{m-1}v_m, v_mv_{m+1}$, where $v_0$ and $v_{m+1}$ are
other nodes, denoted $\prr{C}$ and $\suu{C}$, respectively.  We use
$\head{C}$ to denote $v_1$ and $\tail{C}$ to denote $v_m$. Note that $C$
could be a single node.

We use $[v,w]_{G}$ to denote the chain of $G$ starting from $v$ and
ending at $w$. Let $[v,-]_{G}$ denote the longest chain of $G$
starting from $v$, and $[-,v]_{G}$ the longest chain of $G$ ending
at $v$.  If it is clear from the context which graph is intended, then we
may omit the subscript $G$. Note that the above notation might not be
well-defined for any acyclic graph $G$, but it is so when $G$ is
composed of disjoint chains, which is the case of interest in this
paper. 

\begin{figure}%[p]
\centerline{\input{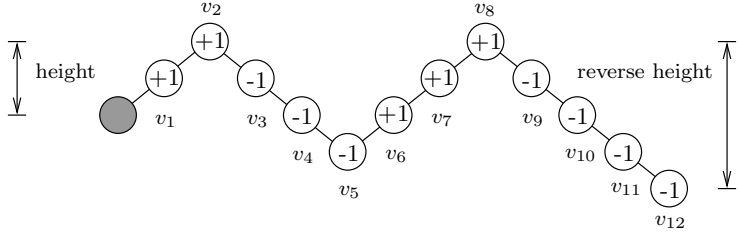}}
%\centerline{\psfig{figure=hump.ps,width=4in,silent=1}}
\caption[]{A hump $H$ of 12 nodes: $v_1,v_2,\ldots,v_{12}$. The cost
  of each node is in the circle. By definition $\cost{H}=-2$,
  $\height{H}=2$, and $\fall{H}=4$. Both of $v_2$ and $v_8$ are peaks of
  $H$, but only $v_2$ is useful.}
\label{hump}
\end{figure}

Suppose $H$ is a chain of $G$ containing a peak $v_\ell$ such that
(1) every node of $H$ preceding $v_\ell$ has nonnegative height in
$H$, and (2) every node of $H$ following $v_\ell$ has height in $H$ at
least the cost of $H$. In this case, we call $H$ a {\em hump}, and we say
$v_\ell$ is a {\em useful peak} of $H$.  This definition is illustrated in
Figure~\ref{hump}\note{Figure~\ref{hump}}.  We say a hump is an {\em $N$-hump} if its
cost is negative, a {\em $P$-hump} if its cost is nonnegative.

We are concerned primarily with graphs $G$ consisting of disjoint
chains $C_1,C_2,\ldots,C_p$.  For convenience, we assume that $G$
contains an {\em initial pseudonode} ($\init$),
preceding all nodes, and a {\em terminal pseudonode} ($\fini$),
following all nodes, each of cost zero. Thus, $\prr{v}$ could be
$\init$ and $\suu{v}$ could be $\fini$.

For the rest of the section we describe the properties of humps in
schedules, mostly adapted from~\cite{Abdel-Wahab:1978:SMM}.

\subsection{Hump Decomposition}
\label{property-sect}

\begin{figure}%[p]
%\centerline{\psfig{figure=chain.ps,width=4in,silent=1}} %
\centerline{\input{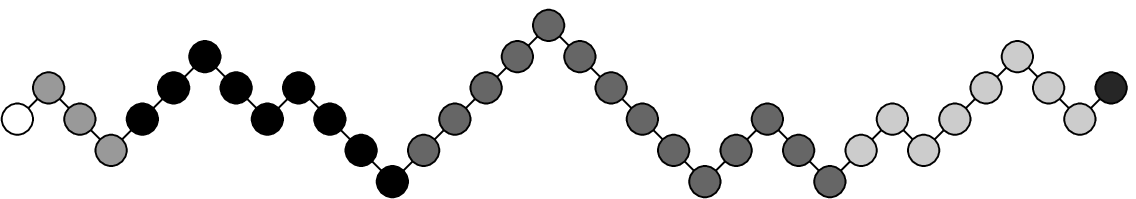}}
\caption[]{A chain decomposed into two $N$-humps and three $P$-humps.}
\label{decompose}
\end{figure}

As part of their scheduling algorithm for series-parallel graphs,
Abdel-Wahab and Kameda~\cite{Abdel-Wahab:1980:SOS} show that in linear
time a sequence of nodes can be decomposed into a set of humps by an
algorithm $\decomp{}$.
%The algorithm $\decomp{}$ is
%shown in Figure~\ref{humpdecomp}\note{Figure~\ref{humpdecomp}}. 
It
takes a chain as input and outputs a set of disjoint subchains such
that every subchain is a hump.  
%The first Repeat-loop produces
%$N$-humps; and the second Repeat-loop produces $P$-humps.  Each loop
%alternates between identifying peaks and valleys.  It is not difficult
%to see that every sequence of nodes between two consecutive valleys is
%a hump.  
The output of $\decomp{C}$ is unique, although the output is not
necessarily the only hump decomposition of $C$.  An example is shown
in Figure~\ref{decompose}\note{Figure~\ref{decompose}}. The chain is
decomposed by $\decomp{}$ into two $N$-humps and three $P$-humps.  For
a chain $C$, we say $H$ is a {\em hump of $C$} if $H\in\decomp{C}$.
It can be proved that $\decomp{}$ has the following properties.

%\begin{figure}%[p]
%\begin{center}
%\fbox{
%\begin{minipage}{5in}
%\begin{center}
%\begin{tabbing}
%\quad\=\quad\=\quad\=\quad\=\quad\=\quad\=\kill
%Function $\decomp{C}$ \+\\
%$S:=\setof{};$\\
%$u:=$ the first valley of $C$;\\
%Repeat\+\\
%$v$\>$:=$ the first peak of $[\prr{C},u]$;\\
%$w$\>$:=$ the first valley of $[\prr{C},v]$;\\
%$S$\>$:=$ $S\cup\setof{[\suu{w},u]};$\\
%$u$\>$:=$ $w;$\-\\
%Until $u=\prr{C}$;\\
%$u:=$ the first valley of $C$;\\
%Repeat\+\\
%$v$\>$:=$ the last peak of $[u,\tail{C}]$;\\
%$w$\>$:=$ the last valley of $[v,\tail{C}]$;\\
%$S$\>$:=$ $S\cup\setof{[\suu{u},w]}$;\\
%$u$\>$:=$ $w;$\-\\
%Until $u=\tail{C}$;\\
%Return $S$;
%\end{tabbing}
%\end{center}
%\newpage
%\end{minipage}
%}
%\end{center}
%\caption[]{The algorithm decomposing a chain into a set of humps.}
%\label{humpdecomp}
%\end{figure}

%\clearpage

\paragraph{Hump-decomposition properties:}
\begin{enumerate}
\item Suppose $H_1, H_2\in\decomp{C}$ and $H_1$ precedes $H_2$ in $C$.
  If $\cost{H_1}\ge0$, then $\cost{H_2}\ge0$ and $\fall{H_1}>\fall{H_2}$.
  If $\cost{H_2}<0$, then $\cost{H_1}<0$ and $\height{H_1}<\height{H_2}$.
\item If $v$ is the first valley of $[u,w]$, then $\decomp{[u,v]}$
(respectively, $\decomp{[\suu{v},w]}$) consists of $N$-humps
(respectively, $P$-humps) only.

\item Let $C$ and $C'$ be two disjoint chains, whose humps are
  respectively $H_1,H_2,\ldots,H_k$ and $H_{k+1},H_{k+2},\ldots,H_\ell$
  in order.  Then, for some $1\le i\le k$ and $k\le j\le \ell$, the humps
  of $CC'$ are 
  \[
     H_1,H_2,\ldots,H_i,(H_{i+1}\cdots H_j),H_{j+1},\ldots,H_\ell
  \]
  in order.
\end{enumerate}
The third property implies that
%\begin{eqnarray*}
\begin{displaymath}
\set{\tail{H}}{H\in\decomp{CC'}}\subseteq
\set{\tail{H}}{H\in\decomp{C}}\cup\set{\tail{H}}{H\in\decomp{C'}}.
\end{displaymath}
%\end{eqnarray*}

It will turn out that once we decompose a chain into humps, we need
not be concerned with the internal structure of these humps. For each
hump $H$ we need only store $\cost{H}$ and $\height{H}$.  Thus, a
chain consisting of $\ell$ humps can be represented by a length-$\ell$
sequence of pairs $(\cost{H},\height{H})$. We call this sequence the
{\em hump representation} of the chain. Using the third
hump-decomposition property, one could straightforwardly derive the
hump representation of $C_1C_2$ from the hump representation of $C_1$
and that of $C_2$.  In particular, if we are given $\decomp{C}$ and
$\decomp{C'}$, then computing $\decomp{CC'}$ takes
$O(|\decomp{C}|+|\decomp{C'}|)$ time.

\begin{figure}%[p]
%\centerline{\psfig{figure=useful.ps,width=4in,silent=1}}
\centerline{\input{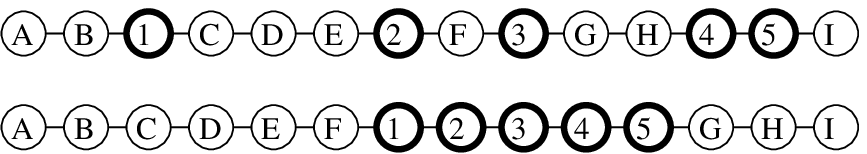}}
\caption[]{The second sequence of nodes is obtained from the first one
by clustering the nodes $1--5$ to node $3$.}
\label{cluster-line}
\end{figure}

\subsection{Hump Clustering}
%%Two lemmas are useful to our results. They are both generalizations
%%of lemmas in [.Kameda 1978.]. 
The following lemma concerns an operation on a schedule called {\em
clustering} the nodes of a hump.  Suppose $H$ is a hump of $G$, and
let $v$ be a useful peak of $H$.  Let $S$ be a schedule of $G$.  If
all the nodes of $H$ are consecutive in $S$, then we say $H$ is {\em
clustered in $S$}.  If every hump of $G$ is clustered in $S$, then we say
the schedule $S$ is {\em clustered}.  If a hump is not clustered in a
schedule, then we can modify the schedule to make it so.  To {\em cluster
the nodes of $H$ to $v$} is to change the positions of nodes of $H$
other than $v$ so that all the nodes of $H$ are consecutive, and the
order among nodes of $H$ is unchanged.  An example is shown in
Figure~\ref{cluster-line}\note{Figure~\ref{cluster-line}}.

\begin{lemma}[See~\cite{Abdel-Wahab:1978:SMM}]
\label{cluster}
%{\rm [. Kameda 1978 .]}
Let $G$ be an acyclic graph with node costs and $H$ be a hump of
$G$. Suppose $S$ is a schedule of $G$. If $T$ is obtained from $S$
by clustering all nodes in $H$ to a useful peak of $H$, then $T$ is a
schedule of $G$ and $\height{T} \le \height{S}$.
\end{lemma}

\begin{figure*}%[p]
%\centerline{\psfig{figure=cluster.ps,width=4.8in,silent=1}}
\centerline{\input{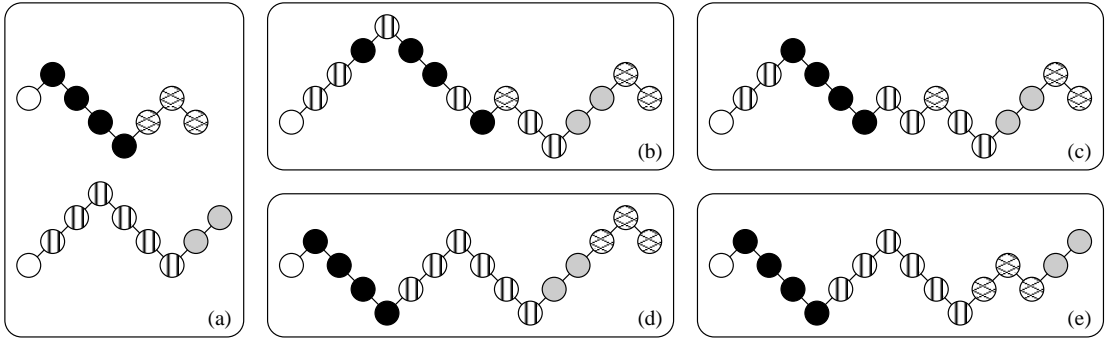}}
\caption[]{(a) A graph $G$ consists of two chains. The first chain
contains an $N$-hump followed by a $P$-hump. The second chain contains
two $P$-humps. (b) A schedule for $G$ of height four. (c) The schedule
obtained from the previous one by clustering the $N$-hump to its
useful peak. (d) A clustered schedule of $G$ of height two. This one
is obtained from the previous schedule by clustering every hump. (e) A
clustered schedule of $G$ with minimum height.}
\label{cluster-fig}
\end{figure*}

An example is shown in
Figure~\ref{cluster-fig}\note{Figure~\ref{cluster-fig}}. The height of
the schedule in Figure~\ref{cluster-fig}(c) is smaller than that of
the schedule in Figure~\ref{cluster-fig}(b).  
%It follows from
%Lemma~\ref{cluster} that there is always a clustered optimal schedule
%of $G$. 
Two clustered schedules of the graph in Figure~\ref{cluster-fig}(a)
are shown in Figures~\ref{cluster-fig}(d) and~\ref{cluster-fig}(e).
It follows from Lemma~\ref{cluster} that there is always an optimal
schedule of $G$ which is clustered.
%We prove the lemma as follows.

\subsection{Standard Order}
A series $S_1 \cdots S_m$ of subsequences of nodes is in {\em standard
order} if it satisfies the following properties.

\paragraph{Standard order properties.}
\begin{itemize}
   \item The series consists of $S_i$'s with negative costs, followed
         by $S_i$'s with nonnegative costs;
   \item The $S_i$'s with negative costs are in nondecreasing order of
         height; and the $S_i$'s with nonnegative costs are in
         nonincreasing order of reverse height.
\end{itemize}

If the humps of a chain are $H_1,H_2,\ldots,H_m$ in order, then the
series $H_1H_2\cdots H_m$ is in standard order by the first
hump-decomposition property.

\begin{lemma}[See~\cite{Abdel-Wahab:1978:SMM}]
\label{exchange}
Let $A$, $B$, $S_1$ and $S_2$ be subsequences of nodes. Suppose
$S=S_1ABS_2$ and $T=S_1BAS_2$. If the series $BA$ is in standard order,
then $\height{S}\ge\height{T}$.
\end{lemma}

For example, the sequence in Figure~\ref{cluster-fig}(d) is a
clustered schedule of the graph in Figure~\ref{cluster-fig}(a). Note
that the series of the last two humps in the schedule is not in
standard order: the reverse height of the first hump (zero) is less
than that of the second hump (one). The schedule in
Figure~\ref{cluster-fig}(e) obtained by exchanging those two
clustered humps has height one less than that of the schedule in
Figure~\ref{cluster-fig}(d).

%\paragraph{Proof of Lemma~\ref{exchange}}
%The original version of this lemma in~\cite{Abdel-Wahab:1978:SMM} restricts
%$A,B$ to be humps $G$. We can prove as follows that the same property
%holds even without these restrictions. Since the heights of nodes in
%$S_1$ and $S_2$ are not changed in $S$ and $T$, it suffices to ensure
%that
%\begin{eqnarray*}
% \height{AB}&=&\maxof{\height{A},\cost{A}+\height{B}}\\
%            &\ge&\maxof{\height{B},\cost{B}+\height{A}}\\
%            &=&\height{BA}.
%\end{eqnarray*}
%\begin{itemize}
%\item If $\cost{A}<0$ and $\cost{B}<0$, since the series $BA$ is in
%         standard order, $\height{A}\ge\height{B}$. Since
%         $\cost{B}<0$, it follows that $\height{A}>\cost{B}+\height{B}$.
%\item If $\cost{A}\ge0$ and $\cost{B}\ge0$, since the series $BA$ is
%         in standard order,
%         $\height{B}-\cost{B}=\fall{B}\ge\fall{A}=\height{A}-\cost{A}$. 
%         Thus $\cost{A}+\height{B}\ge\cost{B}+\height{A}$. Since
%         $\cost{A}\ge0$, $\cost{A}+\height{B}\ge\height{B}$. 
%\item If $\cost{A}\ge0$ and $\cost{B}<0$, then
%    $\height{A}>\cost{B}+\height{A}$ and
%    $\cost{A}+\height{B}\ge\height{B}$.
%\end{itemize}
%Since in all cases each of $\height{B}$ and $\cost{B}+\height{A}$ is
%less than or equal to one of $\height{A}$ and $\cost{A}+\height{B}$,
%the lemma is proved.
%\qed

\subsection{Hump Merging}
A schedule of $G$ is in {\em standard form} if it is clustered and its
series of humps of $G$ is in standard order.  Let $T$ be any schedule
of $G$ in standard form.  Recall that by Lemma~\ref{cluster} there is
always an optimal schedule $S$ of $G$ which is clustered.  The humps
of $G$, while clustered in both $T$ and $S$, may not be in the same
order.  However, any two humps of the same chain of $G$ must be in the
same order in $T$ and in $S$, else either $T$ or $S$ is not a
schedule.  Take two consecutive humps in $S$ that are from different
chains and that are not in the same order as in $T$, and exchange
their positions.  By Lemma~\ref{exchange}, the resulting ordering has
height no more than $S$.  By a series of such exchanges, we eventually
obtain $T$ from $S$.  It follows that the height of $T$ is no more
than that of $S$, and hence that $T$ is optimal.  This argument shows
that every schedule in standard form is an optimal schedule of $G$.

Let $I=\setof{H_1,H_2,\ldots,H_m}$, where the series $H_1H_2\cdots
H_m$ is in standard order. Suppose $\merge{I}$ returns a sequence of
nodes obtained by concatenating all humps in $I$ into standard order.
Namely, $\merge{I}=H_1H_2\cdots H_m$. Assume for uniqueness that
$\merge{}$ breaks ties in some arbitrary but fixed way.  By the above
argument we have the following lemma.

\begin{lemma}[See~\cite{Abdel-Wahab:1978:SMM}]
The output of
\[
   \merge{\bigcup_{1\le i\le p}\decomp{C_i}}
\] 
is an optimal schedule of $G$.
\label{optsched}
\end{lemma}

An example is shown in Figure~\ref{cluster-fig}. Since the schedule in
Figure~\ref{cluster-fig}(e) is clustered and its series of humps is in
standard order, it is an optimal schedule of the graph in
Figure~\ref{cluster-fig}(a).  Abdel-Wahab and
Kameda~\cite{Abdel-Wahab:1978:SMM} showed that $\merge{\bigcup_{1\le
i\le p}\decomp{C_i}}$ can be obtained in $O(n\log p)$ time.  Note that
the output of function $\merge{}$ may not be unique.  Without loss of
generality, however, we may define $\merge{}$ more restrictively as
follows to make its output unique for the same $G$. Suppose $G$ is
composed of disjoint chains, $C_1, C_2, \ldots, C_p$ and
$I=\bigcup_{1\le i\le p}\decomp{C_i}$. Define $\merge{I}=H_1H_2\cdots
H_m$, where $\setof{H_1,H_2,\ldots,H_m}=I$ and the series
$H_1H_2\cdots H_m$ is in standard order. Furthermore, if $H_iH_j$ and
$H_jH_i$ are both in standard order, where $C_{i'}$ contains $H_i$,
$C_{j'}$ contains $H_j$, and $i'<j'$, then $H_i$ precedes $H_j$ in
$\merge{I}$.

\section{Algorithm for Single Pair}
\label{sec:single}
%To detect race conditions between $v$ and $w$, we need to find a valid
%{\em subschedule} containing $v$ and $w$ such that its maximum
%cumulative cost is minimized.  Note that every valid subschedule of
%$G$ is a valid schedule of a prefix subgraph of $G$. A graph $G_0$ is
%a {\em prefix subgraph} of $G$ if (i) there is no arc of $G$ from any
%node of $G-G_0$ to any node of $G_0$; (ii) every arc of $G$ between
%two nodes of $G_0$ is also an arc of $G_0$. Clearly, in the graph of
%interest, i.e., $p$ parallel chains with an augmented arc, every
%prefix subgraph is determined by a cut comprising $p$ cutpoints.

%Let $G$ be a graph composed of disjoint chains, $C_1, C_2, \ldots, C_p$.
%Recall that there are two pseudonodes, $\init$ and $\fini$. The cost of
%each node of $G$ is either $+1$ or $-1$. 
%A subschedule $S$ of $G$ is {\em valid} if $\height{S}=0$.  Let $v$
%and $w$ be two nodes of $G$. In this section we show how to determine
%in linear time whether $v$ could precede $w$ in some valid subschedule
%of $G$.

%\subsection{Notation}
A vector $\cut=(x_1,x_2,\ldots,x_p)$ of $p$ nodes is called a {\em
cut} of $G$ if each $x_i$ is either $\init$ or a node in $C_i$. We
call $x_i$ the $i$-th {\em cutpoint} of $\cut$.  The {\em prefix
subgraph} $G[\cut]$ of $G$ is the subgraph $\bigcup_{1\leq i\leq
p}[-,x_i]$.  Therefore, the problem we address can be reduced to
finding a cut such that the valid schedule of the prefix subgraph
determined by the cut has the minimal maximum cumulative cost.  Let
$h$ be the maximum cumulative cost of the optimal subschedule that
contains $v$ and $w$.  If $h$ is zero, then a valid subschedule exists
(i.e., the optimal valid subschedule.) If $h$ is positive, then there
is no valid subschedule because the maximum cumulative cost of any
valid subschedule is greater than or equal to $h$ and is thus
positive, too.  The rest of the section shows that a best cut can be
found in linear time.

Since we will frequently encounter two cuts that differ at only one
cutpoint, let $\newcut{\cut, i, u}$ denote a cut $\cut'$ with
\begin{displaymath}
   \cut'(\ell)=\left\{
                 \begin{array}{ll}
                    \cut(\ell)&\mbox{if\ $\ell\ne i$};\\
                    u&\mbox{if\ $\ell=i$}.
                 \end{array}
               \right.
\end{displaymath}
A {\em $j$-schedule} of $G[\cut]$ is a schedule of $G[\cut]$ whose
last node is $\cut(j)$. We use $\jopt{\cut}$ to denote the height of
an optimal $j$-schedule of $G[\cut]$. Suppose $\cut(j)\ne\init$. One
can compute $\jopt{\cut}$ for a given $\cut$ as follows. Let
$\cut'=\newcut{\cut,j,\prr{\cut(j)}}$.  Clearly, if $S$ is an optimal
schedule of $G[\cut']$, then $S\cut(j)$ is an optimal $j$-schedule of
$G[\cut]$.
%(However, the other direction is not true.) 
It follows that
\begin{displaymath}
  \jopt{\cut}=\max\setof{\height{G[\cut']},\cost{G[\cut']}+\height{\cut(j)}}. 
\end{displaymath}
Note that $\opt{G[\cut]}$ and $\jopt{\cut}$ are both nonnegative.  We
use $v\chb w$ to signify that there is a valid subschedule of $G$ in
which $v$ precedes $w$. Let $v\notchb w$ signify that $v\chb w$ is not
true. Note that neither $\chb$ nor $\notchb$ is a partial order.

\subsection{Basic Idea}
Every valid subschedule of $G$ is a valid schedule of a prefix subgraph
$G[\cut]$ for some cut $\cut$ of $G$. Therefore, $v\chb w$ if and only if
there is a cut $\cut$ of $G$ such that $G[\cut]$ has a valid schedule in
which $v$ precedes $w$.  Let $h^*$ be the minimum of
$\height{G[\cut]\cup\setof{vw}}$ over all $G[\cut]$'s that contain $v$
and $w$. It follows that $v\chb w$ if and only if $h^*=0$. Hence, the
problem of determining whether $v\chb w$ is reduced to computing the
minimum height of a set of chain graphs each augmented with an
interchain arc.  Clearly, two immediate questions arise. 1) How do we
compute the height of $G[\cut]\cup\setof{vw}$, which is not even
serial-parallel?  2) How do we cope with the fact that there could be
exponential number of prefix subgraphs that contain $v$ and $w$?

Let $v$ and $w$ be contained in two disjoint chains $C_i$ and $C_j$,
respectively.  The following observation will ease the situation.
Suppose $S$ is a subschedule of $G$ containing $w$. Let $S'$ be the
subschedule of $G$ obtained from $S$ by discarding all nodes succeeding
$w$ in $S$. Clearly, $\height{S'}\le\height{S}$. Therefore, without loss
of generality the minimum of $\height{G[\cut]\cup\setof{vw}}$ can be
computed over only cuts $\cut$ with $\cut(j)=w$. Moreover, we can
let $w$ always be the last node of a subschedule by considering only the
minimum-height $j$-schedule of each $G[\cut]$ that contains $v$. The
first question above is no longer an issue.

It turns out that the second question is not an issue, either.  We
will show that in order to obtain the minimum-height of all those
$j$-schedules, it suffices to consider only $O\left(\sqrt{n}\right)$ cuts. In
particular each of those $O(\sqrt{n})$ cuts is uniquely determined by
its $j$-th cutpoint.

\subsection{The Algorithm}
The algorithm takes $v$ and $w$ as inputs. Let $C_i$ contain $v$ and
$C_j$ contain $w$. The algorithm proceeds iteratively with different
cutpoint $\cut(i)$ such that $\cut(i)$ does not precede $v$. In each
iteration the algorithm calls the function $\best{}$ to obtain a
minimum-height $j$-schedule for $G[\cut]$ over all cuts $\cut$ with the
designated cutpoints in $C_i$ and $C_j$. By comparing the heights of
these $j$-schedules with respect to different $\cut(i)$'s, the algorithm
outputs the minimum height of $j$-schedules for $G[\cut]$ over all
$\cut$ such that $\cut(j)=w$ and $\cut(i)$ does not precede $v$. In
Figure~\ref{minheight}\note{Figure~\ref{minheight}} we give the algorithm to
compute $\opt{G[\cut^*]\cup\setof{vw}}$, where $\cut^*$ is a best cut of $G$
corresponding to $vw$.

%%% The function $\best{}$ is the essential part of the algorithm.  {\bf We
%%% need a better explanation than what follows\ldots}.

%%% 
%%% The algorithm takes $v$ and $w$ as inputs. Let $C_i$ contain $v$ and
%%% $C_j$ contain $w$. After fixing $\cut(j)$ at $w$, the algorithm
%%% proceeds iteratively with different $\cut(i)$ in every iteration.
%%% Based on the designated $\cut(i)$ and the fixed $\cut(j)$, the
%%% algorithm obtains a $j$-schedule of minimum height for $G[\cut]$ over
%%% all cuts $\cut'$ such that $\cut'(i)=\cut(i)$ and $\cut'(j)=\cut(j)$.
%%% By comparing the heights of these $j$-schedules with respect to
%%% different $\cut(i)$'s, the algorithm outputs the minimum height of
%%% $j$-schedules for $G[\cut]$ over all cuts $\cut'$ such that
%%% $\cut'(j)=\cut(j)$.
%%% 
%%% If $u$ is a node of $C_k$, by definition $[-,u]=[\head{C_k}, u]$ and
%%% $[u,-]=[u,\tail{C_k}]$.  In Figure~\ref{minheight} we give the
%%% algorithm to compute $\opt{G[\cut^*]\cup\setof{vw}}$, where $\cut^*$
%%% is a best cut of $G$ corresponding to $vw$.  

Function $\best{}$ is the essential part of the algorithm. Based on the
given subset $F$ of $\setof{1,2,\ldots, p}$ and the given cut $\cut$, it
looks for a best cut $\cut^*$ corresponding to $vw$ such that
$\cut^*(k)=\cut(k)$ for every $k\in F$.  (In the case that we are
interested, $F=\setof{i,j}$.)  An optimal $j$-schedule of $G[\cut^*]$ is
then returned. Note that for every $k\not\in F$, $\cut^*(k)$ depends on
a value $s$, which is the maximum of $s_1$ and $s_2$. Each of $s_1$ and
$s_2$ is determined simply by chains with indices in $F$ and their
designated cutpoints.  Namely, the choices of $\cut^*(k)$'s for different
$k\not\in F$ are mutually independent. This is the key to our efficient
algorithm.

In $\best{}$, we do not explicitly specify cutpoints of $\cut^*$.
Instead, we work on hump representation of subchains and every
cutpoint is implicitly specified by an $\tail{H}$ for some hump $H$.
Specifically, Step~1 ensures $\cut^*(k)=\cut(k)$ for every $k\in F,
k\ne j$. Steps~3 and~8 ensure $\cut^*(k)=\tail{H}$, where $H$ is the
highest $N$-hump of all $C_k$ with $\height{H}<s$ and $k\not\in F$.
%that has height less than $s$, for every
%$k\not\in F$. 
Since we are considering $j$-schedules, $\cut^*(j)$ is specified
slightly differently. Although in Step~2 the subchain of $C_j$ is only
up to $\prr{\cut(j)}$, $\cut^*(j)$ is still $\cut(j)$, since
$j$-schedule $S^*\cut(j)$ is returned in Step~10.

\begin{figure*}%[p]
\begin{center}
\fbox{
%\begin{center}
\begin{minipage}[t]{3in}
\begin{tabbing}
Function $\minheight{v,w}$\\
1\quad\=$C_i$\quad\=$:=$ the chain containing $v;$\\
2\>     $C_j$     \>$:=$ the chain containing $w;$\\
3\>     $\cut(j)$ \>$:=w;$\\
4\>     $h^*$     \>$:=\infty;$\\
5\>     $I_0$     \>$:=\setof{v}\cup\decomp{[\suu{v},-]};$\\
6\>     For every $\cut(i) \in \set{\tail{H}}{H\in I_0}$ do\\
7\>\quad\= $S^*$\=$:=\best{j,\setof{i,j},\cut};$\\
8\>\>      $h^*$\>$:=\min\setof{h^*,\height{S^*}};$\\
9\>     Return $h^*;$
\end{tabbing}
\end{minipage}
}
\quad
\fbox{
\begin{minipage}[t]{3in}
\begin{tabbing}%\\ \\
Function $\best{j,F,\cut}$\\
1\quad\=$I$\quad\=$:=\bigcup_{k\in F,k\ne j}\decomp{[-,\cut(k)]};$\\
2\>     $J$\>     $:=\decomp{[-,\prr{\cut(j)}]};$\\
3\>     $K$\>     $:=\bigcup_{k\not\in F}\decomp{C_k};$\\
4\>     $s_1$\>   $:=\max\set{\height{H}}{H\in I\cup J, \cost{H}<0};$\\
5\>     $S^+$\>   $:=\merge{\set{H\in I\cup J}{\cost{H}\ge0}};$\\
6\>     $s_2$\>   $:=\height{S^+\cut(j)};$\\
7\>     $s$\>     $:=\maxof{s_1,s_2};$\\
8\>     $K_s$\>   $:=\set{H\in K}{\height{H}<s, \cost{H}<0};$\\
9\>     $S_s$\>   $:=\merge{I\cup J\cup K_s};$\\
10\>    Return    $S_s\cut(j);$
\end{tabbing}
%\end{center}
\end{minipage}
}
\end{center}
\caption[]{The algorithm for computing $\opt{G[\cut^*]\cup\setof{vw}}$
for a best cut $\cut^*$ of $G$ corresponding to $vw$.}
\label{minheight}
\end{figure*}

\subsection{Correctness}
We answer the following two questions in this subsection:
\begin{enumerate}
\item Why is it sufficient to try for $\cut(i)$ only those nodes in
  $\set{\tail{H}}{H\in I_0}$?
\item Why does $\best{j, F, \cut}$ return an optimal $j$-schedule of
   $G[\cut^*]$ with $\cut^*(k)=\cut(k)$ for every $k \in F$?
\end{enumerate}

\begin{lemma}
Let $\cut$ be a cut of $G$. Suppose $[x,z]$ is a subchain of $G$
containing $\cut(i)$. Let $H$ be the hump of $[x,z]$ containing
$\cut(i)$. Let $y$ be the first valley of $[\prr{H},\cut(i)]$.
If
\[
\cut_1(k)=\left\{
     \begin{array}{ll}
       \cut(k)&\mbox{if $k\ne i$};\\
       \prr{H}&\mbox{if $k=i$ and $y=\prr{H}$};\\
       \tail{H}&\mbox{if $k=i$ and $y\ne\prr{H}$},
     \end{array}
          \right.
\]
then
$\jopt{\cut_1}\le\jopt{\cut}$.
\label{humpboundary}
\end{lemma}
\begin{proof}
Straightforward.
\end{proof}

Note that the $\prr{H}$ in the above lemma is always an $\tail{H'}$ for
some hump $H'$ in $I_0$, which is defined in Step~5 of $\minheight{}$.
Therefore, Lemma~\ref{humpboundary} answers the first question.

By definitions of $I$, $J$, and $K_s$ it is not difficult to see that
the sequence returned by $\best{j,F,\cut}$ is an optimal $j$-schedule of
$G[\cut^*]$ for some cut $\cut^*$ such that $\cut^*(k)=\cut(k)$ for
every $k\in F$. The correctness of $\minheight{}$ thus relies on the
following lemma, which answers the second question.

\begin{lemma}
Let $\cut$ be a cut. Let $F$ be a subset of $\setof{1,2,\ldots,p}$
containing $j$. If $S^*=\best{j,F,\cut}$, then
$\height{S^*}\le\jopt{\cut}$.
\label{fix2cutpoints}
\end{lemma}
The rest of the subsection proves
%Lemma~\ref{humpboundary} and
Lemma~\ref{fix2cutpoints}.
Let $F_\ell=\setof{1,\ldots,\ell-1,\ell+1,\ldots,p}$.  The following
lemma is a special case of Lemma~\ref{fix2cutpoints}, in which $F$ is
composed of $p-1$ numbers.

\begin{lemma}
Let $\cut$ be a cut.  If $S^*=\best{j,F_\ell,\cut}$ for some $\ell\ne
j$, then $\height{S^*}\le\jopt{\cut}$.
\label{bestcut1}
\end{lemma}
\begin{proof}
Define $\cut_1$ by
\[
   \cut_1(k)=\left\{
                   \begin{array}{ll}
                      \cut(k)&\mbox{if $k\ne\ell$};\\
                      \mbox{the first valley of $[-,\cut(\ell)]$}
                            & \mbox{if $k=\ell$}.
                   \end{array}
                \right.
\]
Then it is not difficult to see $\jopt{\cut_1}\le\jopt{\cut}$.  Let
$\cut'$ be the cut with $\height{S^*}=\jopt{\cut'}$, i.e., $S^*$ is a
$j$-schedule of $G[\cut']$.  By definition of $\best{}$, $\cut'$ and
$\cut_1$ could differ only at the $\ell$-th position. Clearly, it
suffices to show $\jopt{\cut'}\le\jopt{\cut_1}$.

Let $w=\cut_1(j)$. Let $L=\decomp{[-,\cut_1(k)]}$. Define
\[
   S=\merge{I\cup J\cup L},
\]
where $I$ and $J$ are defined in Steps~1 and~2 of $\best{}$.  Clearly,
$Sw$ is an optimal $j$-schedule of $G[\cut_1]$. Thus,
$\height{Sw}=\jopt{\cut_1}$.  By choice of $\cut_1(\ell)$, $L$
contains no $P$-hump. Hence, by the uniqueness assumption of
$\merge{}$, we could write $Sw=S_1S^+w$, where $S^+$ is defined in
Step~5 of $\best{}$. We prove $\jopt{\cut'}\le\jopt{\cut_1}$ by 
showing that $\cut'(\ell)$ succeeds $\cut(\ell)$ if and only if
$\jopt{\cut'}\le\height{Sw}$ as follows.

\paragraph{Case 1: $\cut'(\ell)$ succeeds $\cut(\ell)$.}
Since $L$ contains no $P$-hump, each hump of $[-,\cut_1(\ell)]$
appears in $S_1$. Therefore, $S_1S'S^+w$ is a $j$-schedule of
$G[\cut']$, where $S'=[\suu{\cut_1(\ell)},\cut'(\ell)]$.  We show
$\height{S_1S'S^+w}\le\height{S_1S^+w}$.  Now
$\height{S_1S'S^+w}=\max\setof{\height{S_1},\cost{S_1}+\height{S'},\cost{S_1S'}+\height{S^+w}}$.
Clearly,
\begin{equation}
  \height{S_1}\le\height{S_1S^+w}.
  \label{ineq1}
\end{equation}
By definition of $F$, the $K_s$ defined in Step~8 of $\best{}$ is
composed of the $N$-humps of $C_\ell$ that have heights less than $s$.
Therefore, by choice of $\cut'(\ell)$ every hump of $[-,\cut'(\ell)]$
has height less than $s$. It follows from the standard order of humps
in $S'$ that $\height{S'}<s$. By Step~7 of $\best{}$,
$s=\maxof{s_1,s_2}$. If $s=s_2=\height{S^+w}$, as defined in Step~6 of
$\best{}$, then $\cost{S_1}+\height{S'}<\cost{S_1}+\height{S^+w}$. If
$s=s_1=\height{H^*}$, where $H^*$ is a highest $N$-hump in $I\cup J$,
then we could write $S_1=S_2H^*S_3$. It follows that
\begin{eqnarray*}
  \cost{S_1}+\height{S'}&=&\cost{S_2H^*S_3}+\height{S'}\\
                        &<&\cost{S_2}+\height{H^*}\\
                        &\le&\height{S_2H^*}\\
                        &\le&\height{S_1}.
\end{eqnarray*}
Therefore, in either case we have
\begin{equation}
   \cost{S_1}+\height{S'}<\height{S_1S^+w}.
   \label{ineq2}
\end{equation}
By choice of $\cut'(\ell)$, $\cost{S'}<0$. Hence,
\begin{eqnarray}
  \cost{S_1S'}+\height{S^+w}&<&\cost{S_1}+\height{S^+w}\nonumber\\ 
  &\le&\height{S_1S^+w}.  \label{ineq3}
\end{eqnarray}
Combining Equations~(\ref{ineq1}),~(\ref{ineq2}), and~(\ref{ineq3}),
we obtain $\height{S_1S'S^+w}\le\height{Sw}$.

\paragraph{Case 2: $\cut'(\ell)$ precedes $\cut_1(\ell)$.}
Let $S'=[\suu{\cut'(\ell)},\cut_1(\ell)]$. By choice of
$\cut'(\ell)$, it is not difficult to see
\begin{displaymath}
 \decomp{[-,\cut_1(\ell)]}=\decomp{[-,\cut'(\ell)]}\cup\decomp{S'}.
\end{displaymath}
By choice of $\cut_1(\ell)$, $\decomp{S'}$ contains only $N$-humps of
heights no less than $s$. Note that every $N$-hump in $I\cup J$ has
height no more than $s$. By standard form of $S$, we know that $S'$ is
a suffix of $S_1$. Therefore, we could write $Sw=S_2S'S^+w$. Removing
$S'$ from $Sw$, we obtain a $j$-schedule $S_2S^+w$ of $G[\cut']$. We
show $\height{S_2S^+w}\le\height{Sw}$.

Now
$\height{S_2S^+w}=\max\setof{\height{S_2},\cost{S_2}+\height{S^+w}}$. Clearly,
\begin{equation}
  \height{S_2}\le\height{S_2S'S^+w}=\height{Sw}.
  \label{ineq4}
\end{equation}
Since each hump of $S'$ has height no less than $s$,
$\height{S'}\ge s$. 
Hence, $\height{S'S^+w}\ge\height{S'}\ge s\ge s_2=\height{S^+w}$.
It follows that
\begin{eqnarray}
  \cost{S_1}+\height{S^+w}&\le&\cost{S_1}+\height{S'S^+w}\nonumber\\ 
  &\le&\height{Sw}.  \label{ineq5}
\end{eqnarray}
Combining Equations~(\ref{ineq4}) and~(\ref{ineq5}), we obtain
$\height{S_1S^+w}\le\height{Sw}$. 
\end{proof}

Now we are ready to prove Lemma~\ref{fix2cutpoints}.

\paragraph{Proof of Lemma~\ref{fix2cutpoints}}
Recall that $S^*=\best{j,F,\cut}$. Let $\cut'$ be the cut such that
$S^*$ is a $j$-schedule of $G[\cut']$. ($S^*$ is certainly an optimal
$j$-schedule of $G[\cut']$.) We use the algorithm in
Figure~\ref{cuttransform}\note{Figure~\ref{cuttransform}} to prove the
lemma. Procedure $\cuttrans{}$ proceeds with iterations, in which the
value of $\ell$ varies among $\setof{1,\ldots,p}$. If $\ell\not\in F$,
then the value of $\cut(\ell)$ is updated. Since $S$ is an optimal
$j$-schedule of $G[\cut']$, it follows from Lemma~\ref{bestcut1} that
$\jopt{\cut'}\le\jopt{\cut}$ always holds during the while-loop. If we
could show that $\cuttrans{}$ always terminates, then the lemma is
proved.

Let $s^*_1$, $s^*_2$, and $s^*$ be the $s_1$, $s_2$, and $s$ in the
execution of $\best{j,F,\cut}$. Let $s_1$, $s_2$, and $s$ be those in
the execution of $\best{j, F_\ell, \cut}$. The values of $s_1$, $s_2$,
and $s$ change as the while-loop of $\cuttrans{}$ proceeds. We show
that $\cut$ eventually becomes $\cut'$ by arguing that $s$ eventually
becomes $s^*$.

Since $F\subseteq F_\ell$, $s_1\ge s^*_1$ always holds. By definition
of $\best{}$, whenever Step~7 of $\cuttrans{}$ is finished,
$[-,\cut(\ell)]$ contains only $N$-humps. Thus, after the first $p$
iterations of the while-loop, $[-,\cut(\ell)]$ contains no $P$-hump
for every $\ell\not\in F$. Henceforth, $s_2=s^*_2$ and therefore
$s=\maxof{s_1,s_2}\ge\maxof{s^*_1,s^*_2}=s^*$. If $s>s^*$, then
$s=s_1>s^*$. Since $s_1>s^*$, there must be an $N$-hump $H$ in
$\bigcup_{k\not\in F}\decomp{[-,\cut(k)]}$ such that $\height{H}=s_1$.
Since $s=s_1$, in the next iteration when $C_\ell$ contains $H$,
$\cut(\ell)$ will be moved before $H$ by definition of $\best{}$. It
follows that the value of $s$ is nonincreasing and $s$ will become
$s^*$. Once $s=s^*$, in the following $p$ iterations, $\cut(k)$ will
be moved to $\cut'(k)$ for every $k\not\in F$. The algorithm then
terminates.
\qed

\begin{figure}%[p]
\begin{center}
\fbox{
\begin{minipage}{5in}
\begin{center}
\begin{tabbing}
\quad\=\quad\=\quad\=\quad\=\quad\=\quad\=\kill
Procedure $\cuttrans{\cut,\cut^*}$ \\
1 \> $\ell:=0;$\\
2 \> While $\cut^*\ne\cut$ do\\
3 \> \> $\ell:= (\ell \bmod p) + 1;$\\
4 \> \> If $\ell\not\in F$\\
5 \> \> \> $S:=\best{j,F_\ell,\cut};$\\
6 \> \> \> $\cut':=$ the cut such that $S$ is an
           optimal $j$-schedule of $G[\cut']$;\\
7 \> \> \> $\cut := \cut';$
\end{tabbing}
\end{center}
\end{minipage}
}
\end{center}
\caption[]{The algorithm transforms $\cut$ to $\cut^*$. We prove
Lemma~\ref{fix2cutpoints} by showing that this algorithm always
terminates.}
\label{cuttransform}
\end{figure}

\subsection{Implementation}
\label{singleimplementation}
Recall that $\decomp{C}$ runs in time linear in $|C|$, the length of
chain $C$. It follows that the time complexity of Steps~1--5 and
Step~9 of $\minheight{}$ is $O(n)$. Suppose the order of nodes
assigned to $\cut(i)$ in the for-loop is the same as their order in
$C_i$.  In the subsection we focus on implementing $\best{}$ such that
the for-loop runs in time $O(n)$.

\paragraph{Number of Iterations}
The following lemma ensures that the size of $I_0$ is
$O(\sqrt{|C_i|})$. It follows that the number of iterations is
$O(\sqrt{n})$.

\begin{lemma}
Suppose $C$ is a chain with node costs $\pm1$. The number of humps in
$\decomp{C}$ is $O(\sqrt{|C|})$.
\label{rootn}
\end{lemma}
\begin{proof}
Since the costs of nodes are either $+1$ or $-1$, a hump of height
$\ell$ contains at least $\ell$ nodes. For the same reason, a hump of
reverse height $\ell$ contains at least $\ell$ nodes. By the first
hump decomposition property, the heights of the $N$-humps in
$\decomp{C}$ are different, and so are the reverse heights
of the $P$-humps in $\decomp{C}$. If there are $n_1$
$N$-humps and $n_2$ $P$-humps in $\decomp{C}$, then 
$|C|=\Omega(n^2_1+n^2_2)=\Theta((n_1+n_2)^2)$. This proves the lemma.
\end{proof}

\paragraph{Compact Representation of Humps}
For the sake of efficiency, we do not deal with the internal structure
of humps in $\best{}$. It suffices to represent each hump $H$ by a
pair $(\cost{H},\height{H})$ and work on the compact representation of
humps. Therefore, each of the $I$, $J$, and $K$ computed in the first
three steps is a set of pairs. Clearly, each of these three steps
takes $O(n)$ time.  However, the contents of $J$ and $K$ do not change
in different iterations. Thus, Steps~2 and~3 need only be executed
once.

By $F=\setof{i,j}$, we have $I=\decomp{[-,\cut(i)]}$.  Suppose $I_t$
and $\cut_t$ are the $I$ and $\cut$ in the $t$-th iteration for some
$t\ge 2$. By the order of nodes assigned to $\cut(i)$, we need not
recompute $\decomp{[-,\cut_t(i)]}$ from scratch. In the $t$-th
execution of Step~1, $[-,\cut_t(i)]$ is obtained by appending a hump
$[\suu{\cut_{t-1}(i)},\cut_t(i)]$ to $[-,\cut_{t-1}(i)]$. By the
argument following the hump decomposition properties in
\S\ref{property-sect}, the $t$-th execution of Step~1 takes
$O(|I_{t-1}|)$ time. By Lemma~\ref{rootn}, the time complexity of all
executions of Step~1 is $O(n+\sqrt{n}\times\sqrt{n})=O(n)$.

\paragraph{Priority Tree}
To compute $s_1$ efficiently, we resort to a {\em priority tree}, a
complete binary tree with $n+1$ leaves.\footnote{Note that there are
other ways to implement Step 4 to run in linear time. However, the
necessity of priority tree will become clear when we address the
implementation of the all-pairs algorithm.}  Each leaf keeps two
values, {\em count} and {\em maxheight}. The cost of the $(h+1)$-st
leaf is the number of $N$-humps of height $h$ in $I\cup J$. The
maxheight of the $(h+1)$-st leaf is 0 (respectively, $h$), if its
count is zero (respectively, nonzero). The maxheight of an internal
node is the maximum maxheight of its children. It follows that the
maxheight of the root of a priority tree is the correct value of
$s$. The priority tree can be built in time $O(n)$. Whenever a hump is
added to or deleted from $I\cup J$, the priority tree can be updated
in time $O(\log n)$. Since $J$ is fixed, to compute $s_1$ in $t$-th
iteration for every $t\ge 2$, we add humps in $I_t-I_{t-1}$ to $I\cup
J$, remove humps in $I_{t-1}-I_t$ from $I\cup J$, and update the
priority tree. By the third hump decomposition property, we have
\begin{equation}
  \sum_{2\leq t\leq q_i}|I_t-I_{t-1}|+|I_{t-1}-I_t|=O\left(\sqrt{|C_i|}\right),
  \label{changeofI}
\end{equation}
where $q_i$ is the number of humps in $C_i$.
Hence, the time complexity of all executions of Step~4 is
$O(n+\sqrt{n}\times\log n)=O(n)$.

\paragraph{Hump Tree}
To obtain the value of $s_2$, it is not necessary to know the value of
$S^+$. We need only to obtain the height of $S^+\cut(j)$. Similarly,
the actual value of $S_s$ is irrelevant. What we compare in Step~8 of
$\minheight{}$ is the height of $S_s\cut(j)$. We need a data structure
to compute these two heights efficiently.

Let $L$ be a set of humps such that $\height{H}\le n$ and $\fall{H}\le
n$ for every $H\in L$. A {\em hump tree} $T$ for $L$ is a binary tree
composed of two complete binary subtrees. Each subtree has $n+1$
leaves. Let $T_N$ be the left subtree and $T_P$ be the right subtree.
The $(h+1)$-st leaf of $T_N$ associates with the set of $N$-humps of
height $h$ in $L$.  The $(h+1)$-st leaf of $T_P$ associates with the
set of $P$-humps of reverse height $n-h$ in $L$. Let $T_x$ be the
subtree of $T$ rooted at $x$. Let $L_x$ be the set of humps associated
with leaves of $T_x$. Define $\height{T_x}=\height{\merge{L_x}}$ and
$\cost{T_x}=\cost{\merge{L_x}}$.  Clearly, when $L=I\cup J$,
$\height{T_P}=\height{S^+}$ and $\cost{T_P}=\cost{S^+}$. When $L=I\cup
J\cup K_s$, $\height{T}=\height{S_s}$ and $\cost{T}=\cost{S_s}$. The
heights of $S^+\cut(j)$ and $S_s\cut(j)$ can then be computed by
\begin{eqnarray*}
\height{S^+\cut(j)}&=&\maxof{\height{S^+},\cost{S^+}+\height{\cut(j)}};\\
\height{S_s\cut(j)}&=&\maxof{\height{S_s},\cost{S_s}+\height{\cut(j)}}.
\end{eqnarray*}

Let us keep $\height{T_x}$ and $\cost{T_x}$ in $x$ for every node $x$ of
$T$. Therefore, the hump tree $T$ takes $O(n)$ space.  We show how to
compute $\height{T_x}$ and $\cost{T_x}$ for every node $x$ from leaves
to root. When $x$ is a leaf of $T$, the humps in $L_x$ have the same
height if $x$ is in $T_N$, and the same reverse height if $x$ is in
$T_P$. It is not difficult to see that $\cost{T_x}=\sum_{H\in
  L_x}\cost{H}$; and
\[
  \height{T_x}=
    \left\{
       \begin{array}{ll}
          0 & \mbox{if $L_x=\emptyset$};\\
          h & \mbox{if $x$ is the $(h+1)$-st leaf of $T_N$};\\
          \cost{T_x}-h & \mbox{if $x$ is the $(n-h+1)$-st leaf of $T_P$}.
        \end{array}
    \right.
\]

When $x$ is an internal node of $T$, $\height{T_x}$ and $\cost{T_x}$
can be computed by the information kept in the children of $x$.
Suppose $y$ and $z$ are the left and right children of $x$,
respectively. For any $H$ in $L_y$ and $H'$ in $L_z$, by the way we
associate humps with leaves, the series $HH'$ is in standard order.
Hence,
\begin{eqnarray*}
  \height{T_x}&=&\maxof{\height{T_y},\cost{T_y}+\height{T_z}};\\
  \cost{T_x}&=&\cost{T_y}+\cost{T_z}.
\end{eqnarray*}
It follows that the hump tree $T$ for $L$ can be built in time
$O(n+|L|)$.

Once $T$ is built, inserting a hump to $L$ can be done efficiently.
Suppose we insert $H$ to $L$. For the case that $H$ is an $N$-hump, if
$L_x=\emptyset$, then let $h(T_x)=h$; otherwise, add $\cost{H}$ to
$\cost{T_x}$, where $x$ is the $(\height{H}+1)$-st leaf of $T_N$. If
$H$ is a $P$-hump, then we add $\cost{H}$ to both $\cost{T_x}$ and
$\height{T_x}$, where $x$ is the $(n-\fall{H}+1)$-st leaf of $T_P$.
To update $T$, we simply update the internal nodes on the path from
$x$ to the root of $T$. Deleting a hump from $L$ can be done similarly
by replacing every addition with a subtraction. Clearly, both insertion
and deletion take time $O(\log n)$.

To compute the heights of $S^+\cut(j)$ and $S_s\cut(j)$, we need not
maintain a hump tree for $I\cup J$ and another hump tree for $I\cup
J\cup K_s$. Suppose $K^-$ is the set of $N$-humps in $K$, i.e.,
$K^-=\set{H\in K}{\cost{H}<0}$. It suffices to maintain a hump tree
$T$ for $I\cup J\cup K^-$. Since there is no $P$-hump in $K^-$, it is
still true that $\height{T_p}=\height{S^+}$ and
$\cost{T_p}=\cost{S^+}$. Although the hump tree is not for $I\cup
J\cup K_s$, the values of $\height{S_s}$ and $\cost{S_s}$ can be
efficiently obtained by the procedure in
Figure~\ref{remove-range}\note{Figure~\ref{remove-range}}.
Procedure $\removerange{}$ acts as if the $N$-humps of heights no less
than $s$ are removed from the hump tree for $I\cup J\cup K^-$.
Therefore, the resulting $\height{T}$ and $\cost{T}$ are $\height{S_s}$
and $\cost{S_s}$, respectively. Clearly, $\removerange{}$ takes $O(\log
n)$ time.  Since we maintain the hump tree for $I\cup J\cup K^-$ in
every iteration, we use $O(\log n)$ space to keep those modified
information of $T$. After obtaining the information we need, we
restore the hump tree for $I\cup J\cup K^-$ in time $O(\log n)$.

\begin{figure}%[p]
\begin{center}
\fbox{
\begin{minipage}{5in}
\begin{center}
\begin{tabbing}
\quad\=\quad\=\quad\=\quad\=\quad\=\quad\=\kill
Procedure $\removerange{T,s}$ \\
1 \> $y$ := the $s$-th leaf of $T_N$;\\
2 \> While $y$ is not the root of $T_N$ do\\
3 \> \> $x := \mbox{the parent of $y$};$\\
4 \> \> If $y$ is the left child of $x$ then\\
5 \> \> \> $(\height{T_x},\cost{T_x}):=(\height{T_y},\cost{T_y})$;\\
6 \> \> else\\
7 \> \> \> Recompute $\height{T_x}$ and $\cost{T_x}$;\\
8 \> \> $y := x;$\\
9 \> Recompute $\height{T}$ and $\cost{T}$;
\end{tabbing}
\end{center}
\end{minipage}
}
\end{center}
\caption[]{Let $T$ be the hump tree for $I\cup J\cup K^-$. This
procedure acts as if the $N$-humps of heights no less than $s$ are
removed from the hump tree.}
\label{remove-range}
\end{figure}

Let $I_t$ be the $I$ in the $t$-th iteration for any $t\ge1$.  To
obtain the hump tree for $I_t\cup J\cup K^-$ from $I_{t-1}\cup J\cup
K^-$, we need to insert the humps in $I_t-I_{t-1}$ to $T$ and remove
the humps in $I_{t-1}-I_{t}$ from $T$.  Since each insertion and
deletion takes $O(\log n)$ time, it follows from
Equation~(\ref{changeofI}) that the overall time complexity for
obtaining the hump tree from that of previous iteration is
$O(\sqrt{n}\times\log n)$.  Recall that building a hump tree for $L$
takes $O(n+|L|)$ time. Since there are $n$ nodes in $G$, $|I_1\cup
J\cup K^-|=O(n)$. It follows that the time complexity for building a
hump tree for $I_1\cup J\cup K^-$ is $O(n)$.

By the above arguments we implement $\best{}$ such that the overall
time complexity of the while-loop in $\minheight{}$ is $O(n)$.
We therefore have the following theorem.

\begin{theorem}
\label{singlechb}
Suppose $G$ is a graph consisting of $p$ disjoint chains comprising
$n$ nodes, where each node represents either a $P$-operation or a
$V$-operation.  For any two nodes $v$ and $w$ of $G$, one can
determine in $O(n)$ time whether there is a valid subschedule in which
$v$ precedes $w$.
\end{theorem}

\section{Algorithm for All Pairs}
\label{sec:all}
%\subsection{All-Pairs Race-Condition Detection}
%For the purpose of debugging parallel programs, it is important to
%exactly detect all races. Hence, we need to determine the above for all
%pairs of nodes $v$ and $w$.  
In this section we show how to determine the $\chb$ relations for all
pairs of nodes in $G$. The linear-time algorithm for a single pair of
nodes, applied to all $O(n^2)$ pairs, takes time $O(n^3)$.
Fortunately, there is a {\em compact representation} of this
information.  To represent this information, it is sufficient that we
indicate, for each node $v$, and for each chain $C$ not containing
$v$, the first node $w$ in $C$ such that $v$ precedes $w$ in some
valid subschedules. This representation has size $O(np)$, where $n$ is
the number of nodes and $p$ is the number of chains.  The
representation can be used to determine in constant time whether there
is a race between two given operations $v$ and $w$, assuming that the
input $p$ chains are schedulable.\footnote{Since the $p$ chains
represent a trace of a parallel program, the assumption holds. For
arbitrary $p$ chains, one can determine whether they are schedulable
using the algorithm in~\cite{Abdel-Wahab:1978:SMM}.}
%A race exists if
%either operation can precede the other.  
To determine whether $v$ can precede $w$, we obtain the first node in
$w$'s chain that could be preceded by $v$ in some valid subschedules.
If this first node is numbered later than $w$, then $v$ can precede
$w$.  Otherwise, $v$ cannot precede $w$.  We therefore consider the
complexity of constructing such a representation.  Clearly, it can be
constructed by a sequence of calls to the algorithm of Theorem
\ref{singlechb}. We show how to do much better; in fact the time
required by our algorithm is only $O(\log n)$ times the time required
simply to write down the output.

%Recall that $G$ is composed of $p$ disjoint chains,
%$C_1,C_2,\ldots,C_p$, of $n$ nodes. 

\subsection{The Algorithm}

Let $\first{j}{v}$ denote the first node in $C_j$ that could be
preceded by $v$ in some valid subschedule of $G$. The output of the
all-pairs algorithm is thus the value of $\first{j}{v}$ for every node
$v$ and $1\le j\le p$. Note that $\first{j}{v}$ could be $\fini$,
which means that none of nodes in $C_j$ can be preceded by $v$ in any
valid subschedule of $G$.

Let us describe first the procedure $\chainpair{i,j}$ which computes
$\first{j}{v}$ for every $v\in C_i$. The all-pairs algorithm simply
calls $\chainpair{i,j}$ for every $1\le i, j\le p$.  For convenience, let
$\su{j}{w}=\suu{w}$ for every $w\in C_j$ and let
$\su{j}{\init}=\head{C_j}$.  Procedure $\chainpair{i,j}$ is shown in
Figure~\ref{chainpair}\note{Figure~\ref{chainpair}}.  The algorithm starts
with letting $v$ be $\tail{C_i}$ and letting $w$ be $\tail{C_j}$. The
repeat-loop proceeds by replacing $w$ with $\prr{w}$. Once $\minheight{v,w}$
is not zero, the algorithm reports $\su{j}{w}$ as $\first{j}{v}$. After
replacing $v$ with $\prr{v}$, the repeat-loop continues the same procedure
to search for new $\first{j}{w}$.

\begin{figure}%[p]
\begin{center}
\fbox{
\begin{minipage}{5in}
\begin{center}
\begin{tabbing}
\quad\=\quad\=\quad\=\quad\=\quad\=\quad\=\kill
Procedure $\chainpair{i,j}$ \\
 1\>     $(v, w) := (\tail{C_i}, \tail{C_j})$;\\
 2\>     Repeat\\
 3\>        \>If $w=\init$ \=then \=$h := 1$;\\
 4\>        \>             \>else \>$h := \minheight{v,w}$;\\
 5\>        \>If $h>0$ \>then \=$\first{j}{v} :=\su{j}{w}$;\\
 6\>        \>         \>     \>          $v := \prr{v}$;\\
 7\>        \>         \>else\>$w := \prr{w}$;\\
 8\>     Until $v = \init$;
\end{tabbing}
\end{center}
\end{minipage}
}
\end{center}
\caption[]{The algorithm that computes $\first{j}{v}$ for every $v \in C_i$.}
\label{chainpair}
\end{figure}

\subsection{Correctness}
By induction on $v$ we show that $\chainpair{i,j}$ correctly computes
$\first{j}{v}$ for every $v \in C_i$.

When $v=\tail{C_i}$, procedure $\chainpair{i,j}$ keeps replacing $w$
with $\pr{j}{w}$ until $w = \init$ or $\minheight{v,w}> 0$.  If
$w=\init$, then $\opt{G\cup\setof{vw'}}=0$ for every $w' \in C_j$.  Thus,
$\first{j}{v}=\su{j}{\init}=\su{j}{w}=\head{C_j}$ is correct. If
$\minheight{v,w}>0$, then $v\notchb w$. It follows that $v\notchb w'$ for
every $w'$ precedes $w$ in $C_j$. Since $\minheight{v,\su{j}{w}}=0$,
$v\chb\su{j}{w}$.  Therefore, $\su{j}{w}$ is the correct value of
$\first{j}{v}$. This confirms the induction basis.

Suppose the procedure $\chainpair{i,j}$ correctly reports $\su{j}{w}$ as
the value of $\first{j}{\su{i}{v}}$ in a certain iteration of the
repeat-loop.  We need to show that in the remaining iterations
$\first{j}{v}$ will also be correctly computed. Since
$\su{i}{v}\chb\su{j}{w}$, $v\chb\su{j}{w}$. It follows that $v\chb w'$
(and thus $\minheight{v,w'}=0$) for every $w'$ succeeding $w$ in $C_j$.
In other words, to locate the first node in $C_j$ that could be preceded
by $v$, it suffices to start testing from $w$.  For the same reason as
above, $\chainpair{i,j}$ reports the correct value of $\first{j}{w}$.
The correctness is therefore ensured.

\subsection{Implementation}
We show in this subsection how to implement $\chainpair{i,j}$ to run
in time $O((|C_i|+|C_j|)\log n)$. It then follows that the time
complexity of the all-pairs algorithm is $O(np\log n)$.

Suppose each time before we call $\chainpair{i,j}$, we have the hump
tree for $I\cup J\cup K^-$, where
\begin{eqnarray*}
 I&=&\decomp{C_i};\\
 J&=&\decomp{[-,\prr{\tail{C_j}}]};\\
 K^-&=&\set{H\in\bigcup_{{\scriptstyle1\le k\le p}\atop{\scriptstyle k\ne i,j}}
            \decomp{C_k}}{\cost{H}<0}.
\end{eqnarray*}
It follows from \S\ref{singleimplementation} that the first
call to $\minheight{v,w}$ can be computed in time $O(\log n)$, since
only one $\cut(i)$ need be considered.  In each of the remaining
iterations of the repeat-loop, we either replace $v$ with $\prr{v}$ or
replace $w$ with $\prr{w}$. The remaining lemma guarantees that to
compute each of the following $\minheight{v,w}$, we need only try $v$ as
the cutpoint of $C_i$.

\begin{lemma}
Consider any iteration of the repeat-loop in $\chainpair{i,j}$. When
the algorithm computes $h=\minheight{v,w}$, $v$ is the only cutpoint
of $C_i$ that could make $h$ zero.
\label{onlycutpoint}
\end{lemma}
\begin{proof}
By definition of $\chainpair{}$, when computing $\minheight{v,w}$,
$\first{j}{\su{i}{v}}$ always succeeds $w$ in $C_j$.  Assume for a
contradiction that $u$ is a node succeeding $v$ in $C_i$ such that
there is a cut $\cut$ of $G$ where $\cut(i)=u$, $\cut(j)=w$, and
$\jopt{\cut}=0$. It follows that $u\chb w$ and thus $\su{i}{v}\chb w$.
This contradicts the fact that $\first{j}{\su{i}{v}}$ succeeds $w$ in
$C_j$.
\end{proof}

\begin{theorem}
  \label{allchb}
  Suppose $G$ is as in Theorem \ref{singlechb}.  The compact
  representation of the relation ``$v$ precedes $w$ in some valid
  subschedules'' can be constructed in $O(np\log n)$ time and $O(n)$
  space.
\end{theorem}
\begin{proof}
Note that in each iteration of the repeat-loop, either $v$ or $w$ is
moved by one position. Since the costs of $v$ and $w$ are $\pm1$, by
the first hump decomposition property the number of humps updated in
$I\cup J\cup K^-$ between two consecutive iterations is a constant.
Thus, each execution of $\minheight{v,w}$ takes only time $O(\log n)$.
Since the number of iterations of the repeat-loop is $O(|C_i|+|C_j|)$,
each execution of $\chainpair{i,j}$ takes time
\begin{equation}
  O((|C_i|+|C_j|)\times\log n).
  \label{time1}
\end{equation}
It remains to show how to efficiently build the hump tree for each
execution of $\chainpair{i,j}$.

The very first hump tree can be constructed in time
\begin{equation}
  O(n).
  \label{time2}
\end{equation}
Consider the moment when $\chainpair{i,j}$ is just finished and the
all-pairs algorithm is about to call $\chainpair{i_1, j_1}$.  Since
all humps in $I\cup J$ have been deleted during the execution of
$\chainpair{i,j}$, the current $T$ is the hump tree for the $N$-humps
in $\bigcup_{1\le k\le p; k\ne i,j}\decomp{C_k}$.  In order to obtain
the hump tree for $\chainpair{i_1, j_1}$, we have to add the $N$-humps
in $\decomp{C_i}\cup\decomp{C_j}$, delete the $N$-humps in
$\decomp{C_{j_1}}$ from $T$, and then insert the humps in
\begin{displaymath}
\set{H\in\decomp{C_{i_1}}}{\cost{H}\ge0}\cup\decomp{[-,\prr{\tail{C_{j_1}}}]}
\end{displaymath}
to $T$. The hump decomposition can be done in time
\begin{equation}
  O(|C_i|+|C_j|+|C_{i_1}|+|C_{j_1}|).
  \label{time3}
\end{equation}
The insertion and deletion of humps can be done in time
\begin{equation}
  O\left(\left(\sqrt{|C_i|}+\sqrt{|C_j|}+\sqrt{|C_{i_1}|}+\sqrt{|C_{j_1}|}\right)\times\log n\right).
  \label{time4}
\end{equation}
By (\ref{time1}),~(\ref{time2}),~(\ref{time3}), and~(\ref{time4}), the
overall time complexity of the all-pairs algorithm is
\begin{displaymath}
O(n)+\sum_{1\le i,j\le p}\left(O(|C_i|+|C_j|) +
O\left(\sqrt{|C_i|}+\sqrt{|C_j|}\right) \times\log n +
O(|C_i|+|C_j|)\times\log n\right), 
\end{displaymath}
which is $O(np\log n)$.  
%Theorem~\ref{allchb} is proved.
\end{proof}

\section{NP-completeness}
\label{sec:2semaphores}

In this section we sketch the proof for the following theorem.
\begin{theorem}
\label{2semaphores}
The race-condition detection problem for a parallel program that uses
more than one semaphore is NP-complete.
\end{theorem}
%the NP-complete proof for determining
%whether $v\chb w$ for chain graphs of operations on more than one
%semaphore.  
The proof is by reduction from the NP-complete
uniform-cost SMMCC problem, where the node costs are restricted to
$\pm1$~\cite{Garey:1979:CIG}. The reduction has three steps. Given a SMMCC
problem for a uniform-cost graph $G_0$ of $n$ nodes, we construct
$O(\log n)$ chain graphs with $n+2$ semaphores. The first step of the
reduction shows that the SMMCC problem for $G_0$ can be reduced to
determining whether each of those $O(\log n)$ chain graphs has a valid
schedule.  The second step shows that each of those $O(\log n)$ chain
graphs can be {\em simulated} by a chain graph with only two
semaphores. In other words, the simulated chain graph has a valid
schedule if and only if the simulating chain graph has a valid
schedule.  The last step shows that the simulating chain graph has a
valid schedule if and only if $v\chb w$, for some $v$ and $w$, in the
same chain graph.  We elaborate the details of the reduction in the
appendix.

\section*{Acknowledgments}
We thank the anonymous referees for their helpful remarks that
significantly improved the presentation of the paper.

%
%\section*{References}
%\label{section:refs}\frenchspacing\indent
%.[]
\bibliographystyle{abbrv}
\bibliography{race}

\appendix
\section{Appendix}
%\subsection{Definition and Notation}
Let $G$ be a chain graph. Each node of $G$ is an operation on a
semaphore. An operation on semaphore $S$ is either $+S$, incrementing
the value of $S$ by one, or $-S$, decrementing the value of $S$ by
one.  A subschedule of $G$ is {\em valid} if the value of each
semaphore is always nonpositive during the execution of the
subschedule.  Let $v$ and $w$ be two nodes of $G$.  If there exists a
subschedule of $G$ in which $v$ precedes $w$, then we say $v\chb w$.
Clearly, determining whether $v\chb w$ is in NP.  If $G$ is allowed to
use more than one semaphore, then we prove the NP-hardness by a
three-step reduction from the uniform-cost SMMCC problem.

\subsection{First Step}
Let $G_0$ be an acyclic directed graph of $n$ nodes, $v_1,v_2,\ldots,
v_n$. The cost of each node is either $+1$ or $-1$.  Suppose we would
like to know whether $\height{G_0}\leq\ell$. We construct a chain graph
$G_1$ composed of $2n+2$ chains of operations on $n+2$ semaphores, and
argue that $G_1$ has a valid schedule if and only if
$\height{G_0}\leq\ell$.  Note that $0\leq\height{G_0}\leq n$. Therefore,
$\height{G_0}$ can be obtained by $O(\log n)$ queries of whether a chain
graph of $n+2$ semaphores has a valid schedule.

Let $n^+$ be the number of nodes with positive costs. Let $n^-$ be the
number of nodes with negative costs. Clearly, $n^+-n^-$ is the sum of
node costs of $G_0$. Let $d_i$ be the number of outgoing arcs of $G_0$
from $v_i$.  The $n+2$ semaphores for $G_1$ are
$S_1,S_2,\ldots,S_n,S_\alpha,S_\beta$.
%To distinguish the last two semaphores, we
%also write $S_{\alpha}=S_{n+1}$ and $S_{\beta}=S_{n+2}$.  
Let the $2n+2$ chains of $G_1$ be $C_1,\ldots, C_{n+1}$, and
$C'_1,\ldots,C'_{n+1}$, all initially empty.  We construct $G_1$ from
$G_0$ by the procedure $\construct{}$ in
Figure~\ref{fig:construct}\note{Figure~\ref{fig:construct}}, which
runs in polynomial time. Without loss of generality we can assume that
$\ell-n^++n^-$, the number in the second-to-last statement of the
procedure \fname{Construct}, is nonnegative, since otherwise
$\height{G_0}>\ell$ is immediately concluded.

\begin{figure}%[p]
  \begin{center}
    \fbox{
      \begin{minipage}{4in}
        \begin{tabbing}
          \quad\quad\=\quad\=\quad\=\quad\=\quad\=\quad\=\quad\=\kill
          $\construct{G_0}$\\
          1\>For $i:=1$ to $n$ do\\
          2\>\>For $j:=1$ to $n$ do\\
          3\>\>\>If $v_jv_i$ is an arc of $G_0$ then\\
          4\>\>\>\>Append a $+S_j$ to $C_i$.\\
          5\>\>If the cost of $v_i$ is $+1$ then\\
          6\>\>\>Append a $+S_{\alpha}$ to $C_i$.\\
          7\>\>else (i.e., the cost of $v_i$ is $-1$)\\
          8\>\>\>Append a $-S_{\alpha}$ to $C_i$.\\
          9\>\>\>Append a $+S_{\alpha}$ and $-S_{\alpha}$ to $C'_i$.\\
          10\>\>Append $d_i$ copies of $-S_i$ to $C_i$.\\
          11\>\>Append a $-S_{\beta}$ to $C_i$.\\
          12\>Append $n$ copies of $+S_{\beta}$ to $C_{n+1}$.\\
          13\>Append $\ell-n^++n^-$ copies of $+S_{\alpha}$ to $C_{n+1}$.\\
          14\>Append $\ell$ copies of $-S_{\alpha}$ to $C'_{n+1}$.
        \end{tabbing}
      \end{minipage}
      }
  \end{center}
  \caption{The procedure constructs a chain graph $G_1$ such that $G_1$
    has a valid schedule if and only if $\height{G_0}\leq\ell$.}
  \label{fig:construct}
\end{figure}

\begin{figure*}%[p]
  
%  \begin{center}
%    \leavevmode
%    \begin{tabular}{c}
%      \ovalnode{a}{$v_1:+1$}\\\\
%      \ovalnode{b}{$v_2:+1$}\quad\ovalnode{c}{$v_3:+1$}\\\\
%      \ovalnode{e}{$v_4:-1$}\quad\ovalnode{d}{$v_5:-1$}
%    \end{tabular}
%    \ncline{->}{a}{b}
%    \ncline{->}{a}{c}
%    \ncline{->}{b}{e}
%    \ncline{->}{c}{d}
%    \ncline{->}{c}{e}
%    \ncline{->}{d}{e}
%  \end{center}
%  \vspace{0.5in}
  \begin{center}
    \input{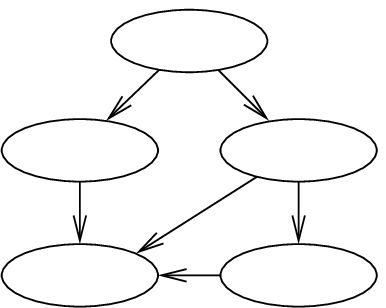}\qquad
    \begin{tabular}[b]{|r|r|r|r|r||r||r|r||r|}
      $C_1$ &$C_2$ &$C_3$ &$C_4$ &$C_5$ &$C_6$ &$C'_4$&$C'_5$&$C'_6$\\
      \hline                                                        
            &      &      &$+S_2$&      &$+S_{\beta}$&$+S_{\alpha}$&$+S_{\alpha}$&$-S_{\alpha}$\\
            &      &      &$+S_3$&      &$+S_{\beta}$&$-S_{\alpha}$&$-S_{\alpha}$&$-S_{\alpha}$\\
            &$+S_1$&$+S_1$&$+S_5$&$+S_3$&$+S_{\beta}$&      &      &      \\
            &      &      &      &      &$+S_{\beta}$&      &      &      \\
      $+S_{\alpha}$&$+S_{\alpha}$&$+S_{\alpha}$&$-S_{\alpha}$&$-S_{\alpha}$&$+S_{\beta}$&      &      &      \\
            &      &      &      &      &      &      &      &      \\
      $-S_1$&$-S_2$&$-S_3$&      &$-S_5$&      &      &      &      \\
      $-S_1$&      &$-S_3$&      &      &      &      &      &      \\
            &      &      &      &      &      &      &      &      \\
      $-S_{\beta}$&$-S_{\beta}$&$-S_{\beta}$&$-S_{\beta}$&$-S_{\beta}$&$+S_{\alpha}$&      &      &
    \end{tabular}
    \caption[]{An example for the first step of the reduction. Suppose
      we would like to determine whether $\height{G_0}\leq2$, where
      $G_0$ is the graph on top. We then construct, by
      \fname{Construct}, the chain graph $G_1$ at bottom. Note that there
      are one $+S_{\alpha}$ at the end of $C_6$ and two $-S_{\alpha}$ in
      $C'_6$, according to the last two statements of \fname{Construct}. It
      follows from Lemmas~\ref{lemma:np-1}(1) and~\ref{lemma:np-2} that
      that exists a valid schedule of the chains at bottom if and only if
      the height of the graph on top is at most two.}
    \label{fig:np-complete1}
  \end{center}
\end{figure*}

An example is shown in
Figure~\ref{fig:np-complete1}\note{Figure~\ref{fig:np-complete1}}.
The intuition is as follows. The (only) operation for $S_{\alpha}$ in
$C_i$ corresponds to $v_i$, where the ``sign'' of $S_{\alpha}$
reflects the cost of $v_i$.  We use the first $n$ semaphores,
$S_1,\ldots,S_n$, to enforce the execution of these $n$ operations for
$S_{\alpha}$ to obey the precedence constraints imposed by $G_0$. In
Figure~\ref{fig:np-complete1}, for instance, in order to reach the
$-S_{\alpha}$ in $C_4$, we have to unlock the $+S_2$ (and $+S_3$,
$+S_5$) in the same chain first.  Since the only $-S_2$ is after the
$+S_{\alpha}$ in $C_2$, we know the $+S_{\alpha}$ in $C_2$ must be
executed before the $-S_{\alpha}$ in $C_4$.

The $-S_{\beta}$'s at the end of $C_1,\ldots,C_n$ are to ensure that as
long as the last $+S_{\beta}$ in $C_{n+1}$ is executed, all operations
in $C_1,\ldots, C_n$ are already executed. The function of those $\ell$
copies of $-S_{\alpha}$ in $C'_{n+1}$ is clear: The larger $\ell$, the
easier for $G_1$ to have a valid schedule.  The purpose of the
$+S_{\alpha}, -S_{\alpha}$ pairs in $C'_1,\ldots, C'_n$ and those
$\ell-n^++n^-$ copies of $+S_{\alpha}$'s at the end of $C_{n+1}$ will
become clear as we proceed.  Basically they are used to ensure that
$G_1$ has some kind of ``pairwise'' schedule, as long as $G_1$ has a
valid schedule. One can verify that there are the same number of
$+S_i$'s and $-S_i$'s in $G_1$, for each $1\leq i\leq n+2$.

For the rest of the subsection, we prove that $\height{G_0}\leq\ell$
if and only if $G_1$ has a valid schedule. An implication of the
following proofs is that $G_1$ has a valid schedule if and only if it
has a valid schedule executable by some procedure \fname{Pairwise},
which will be given in the proofs.
\begin{lemma}
  \label{lemma:np-1}
\begin{enumerate} 
\item If $G_1$ has a valid subschedule containing the last $+S_{\alpha}$
  of $C_{n+1}$, then $\height{G_0}\leq\ell$.
\item
  If $G_1$ has a valid schedule, then $\height{G_0}\leq\ell$.
\end{enumerate}

\end{lemma}
\begin{proof}
  Clearly, it suffices to prove the first statement, since the second
  statement follows immediately from the first statement.
  
  Let $X$ be a valid subschedule of $G_1$ as described in the lemma. We
  show $\height{G_0}\leq\ell$.  Let $O_i$ be the operation of
  $S_{\alpha}$ in $C_i$. Since $X$ is valid and contains the last
  $+S_{\alpha}$ of $C_{n+1}$, $X$ must contain all the operations in
  $C_1,\ldots,C_n$.  Therefore, every $O_i$, $1\leq i\leq n$, is in $X$.

  Suppose the order of those $O_i$'s in $X$ is $O_{k_1},O_{k_2},\ldots,
  O_{k_n}$. By the definition of \fname{Construct}, if $v_j$ is
  reachable from $v_i$ in $G_0$, then $O_j$ does not precede $O_i$ in
  $X$. It follows that the sequence $Y=v_{k_1}v_{k_2}\cdots v_{k_n}$ is
  a schedule of $G_0$.  Therefore, it suffices to show 
  $\height{Y}\leq\ell$.

  Assume $\height{Y}>\ell$ for a contradiction. If
  we count only those $O_i$'s as the operations for $S_{\alpha}$ in $X$, then
  the maximum value of $S_{\alpha}$ would be greater than $\ell$ during
  the execution of $X$. Note that there are $\ell+n^-$ other
  $-S_{\alpha}$'s in $C'_1,\ldots,C'_{n+1}$, which are the only hope for
  bringing the maximum value of $S_{\alpha}$ down to zero.  By the
  construction of $C'_1,\ldots, C'_n$, however, we know $n^-$ of those
  $-S_{\alpha}$'s have to be preceded in $X$ by $n$ other
  $+S_{\alpha}$'s. It follows that even if we count all operations for
  $S_{\alpha}$ together, the maximum value of $S_{\alpha}$ would be
  greater than zero during the execution of $X$.  This contradicts the
  fact that $X$ is a valid schedule of $G_1$.
\end{proof}

\begin{lemma}
  \label{lemma:np-2}
  If $\height{G_0}\leq\ell$, then $G_1$ has a valid schedule.
\end{lemma}
\begin{proof}
  Let $Y=v_{k_1}v_{k_2}\cdots v_{k_n}$ be a schedule of $G_0$ with
  $\height{Y}\leq\ell$.  Let $m_i$ be the sum of costs of
  $v_{k_1},\ldots,v_{k_i}$. Clearly, $m_n=n^+-n^-$, which is the sum of
  node costs of $G_0$.  Since $\height{Y}\leq\ell$, we know that
  $m_i\leq\ell$ for every $1\leq i\leq n$. We claim that $G_1$ can be
  executed by the procedure \fname{Pairwise} in
  Figure~\ref{fig:pairwise}\note{Figure~\ref{fig:pairwise}}.
\begin{figure}%[p]
  \begin{center}
    \fbox{
    \begin{minipage}{5in}
      \begin{tabbing}
        \qquad\=\quad\=\quad\=\quad\=\quad\=\quad\=\quad\=\kill
        Procedure \fname{Pairwise}\\
        1\>For $k:=k_1,k_2,\ldots,k_n$ do\\
        2\>\>For $j:=1$ to $n$ do\\
        3\>\>\>If $v_j v_k$ is an arc of $G_0$ then\\
        4\>\>\>\>Execute a $-S_k$ in $C_j$.\\
        5\>\>\>\>Execute the $+S_k$ in $C_k$.\\
        6\>\>If $O_k = +S_{\alpha}$ then\\
        7\>\>\>Execute one of the $-S_{\alpha}$'s\\
        8\>\>\>\>in $C'_1,C'_2,\ldots,C'_{n+1}$.\\
        9\>\>\>Execute the $+S_{\alpha}$ in $C_k$.\\
        10\>\>else (i.e., $O_k = -S_{\alpha}$)\\
        11\>\>\>Execute the $-S_{\alpha}$ in $C_k$.\\
        12\>\>\>Execute the $+S_{\alpha}$ in $C'_k$.\\
        13\>For $i:=1$ to $n$ do\\
        14\>\>Execute the $-S_{\beta}$ in $C_i$.\\
        15\>\>Execute a $+S_{\beta}$ in $C_{n+1}$.\\
        16\>For $i:=1$ to $\ell-m_n$ do\\
        17\>\>Execute a $-S_{\alpha}$ in $C'_1,\ldots,C'_{n+1}$.\\
        18\>\>Execute a $+S_{\alpha}$ in $C_{n+1}$.
      \end{tabbing}
    \end{minipage}
    }
  \end{center}
  \caption{Procedure \fname{Pairwise}.}
  \label{fig:pairwise}
\end{figure}

Note that in the schedule of $G_1$ executed by \fname{Pairwise}, each
operation $-S_i$ is immediately followed by an operation $+S_i$.  Not
every chain graph has such a ``pairwise'' schedule, however, we show
that $G_1$ does.  We first show that the first for-loop of
\fname{Pairwise} can be finished for $G_1$. Specifically, suppose the
following claim hold:
\begin{quote}
\label{lemma:pairwise-1} {\bf Claim}
  For each $1\leq i\leq n$, the $i$-th iteration of the first for-loop
  of \fname{Pairwise} can be executed for $G_1$. Furthermore, after
  executing the $i$-th iteration,
  \begin{itemize}
  \item the remaining operations in $C_{k_i}$ are $d_{k_i}$ copies of
    $-S_{k_i}$'s followed by a $+S_{\beta}$; and
  \item there are $\ell-m_i$ copies of $-S_{\alpha}$'s available in
    $C'_1,\ldots,C'_{n+1}$.
  \end{itemize}
\end{quote}
It is then not hard to see that after the execution of the first
for-loop of \fname{Pairwise}, the remaining operation in each $C_i$ is a
$-S_{\beta}$. Therefore, the second for-loop of \fname{Pairwise} can be
finished, since there are $n$ copies of $+S_{\beta}$'s available in
$C_{n+1}$.  

By Lemma~\ref{lemma:pairwise-1}, we know that after executing the first
For-loop, the number of $-S_{\alpha}$'s in $C'_1,\ldots,C'_{n+1}$ is
$\ell-m_n$, which is equal to the number of $+S_{\alpha}$'s at the end
of $C_{n+1}$. Therefore, the last for-loop of \fname{Pairwise} can be
finished. The lemma is proved.

It remains to prove the above claim by induction on $i$. For
convenience we abbreviate $k_i$ to $k$ for the rest of the proof.
When $i=1$, we know $v_k$ does not have any incoming arcs from other
nodes. Therefore, the for-loop with index $j$ in the first iteration does
not execute any operation. We then consider the if-statement.
\begin{itemize}
\item If $O_k=-S_{\alpha}$, then $\cost{v_k}=-1$, and thus $m_1=-1$.
  There is a $+S_{\alpha}$ in $C'_k$ by the definition of
  \fname{Construct}. We can execute the else-part of the if-statement
  without problem. Since the second operation in $C'_k$ is a
  $-S_{\alpha}$, these two steps increase the number of $-S_{\alpha}$'s
  available in $C'_1,\ldots,C'_{n+1}$ by one.
\item If $O_k=+S_{\alpha}$, then $\cost{v_k}=1$, and thus $m_1=1$. Since
  $v_k$ is the first node in $Y$, $\height{Y}$ is at least one, and thus
  $\ell\geq 1$. We can therefore execute the then-part of the
  if-statement without problem.  The number of $-S_{\alpha}$'s available
  in $C'_1,\ldots,C'_{n+1}$ is decreased by one.
\end{itemize}
Clearly, after executing the first iteration, in which the only executed
operation in $C_k$ is $O_k$, the remaining operations in $C_k$ are
exactly as that described in the claim.  Note that before executing the
first iteration, the number of available $-S_{\alpha}$'s is $\ell$ by
the definition of \fname{Construct}.  Therefore, after executing the
first iteration, the number of available $-S_{\alpha}$'s is exactly
$\ell-m_1$. This confirms the inductive basis.

Let $i'$ be an integer with $1<i'\leq n$. Assume that the claim holds
for every $1\leq i<i'$. We show it holds for $i=i'$.  Consider the
$i$-th iteration. Note that for every $j$ such that $v_jv_k$ is an arc
of $G_0$, $O_j$ must have been executed.  By the inductive hypothesis
we know those $d_j$ copies of $-S_j$'s are already available before
executing the $i$-th iteration. Therefore, the for-loop with index $j$
will proceed without problem, since there are exactly $d_j$ copies of
$+S_j$'s in $G_1$ by the definition of \fname{Construct}.  We then
consider the if-statement.
\begin{itemize}
\item If $O_k=-S_{\alpha}$, then $m_i=m_{i-1}-1$. We know there is a
$+S_{\alpha}$ in $C'_k$.  Thus, the else-part can proceed without
problem. Since the second operation in $C'_k$ is a $-S_{\alpha}$,
these two steps increase the number of available $-S_{\alpha}$'s in
$C'_1,\ldots,C'_{n+1}$ by one.
\item If $O_k=+S_{\alpha}$, then $m_i=m_{i-1}+1$. The inductive hypothesis
says that the number of $-S_{\alpha}$'s available in
$C'_1,\ldots,C'_{n+1}$ is $\ell-m_{i-1}$ before executing the $i$-th
iteration.  That number is at least one since
$\ell-m_{i-1}-1=\ell-m_i\geq 0$. Therefore, the then-part of the
if-statement can be executed without problem. The number of available
$-S_{\alpha}$'s in $C'_1,\ldots,C'_{n+1}$ is decreased by one.
\end{itemize}
Therefore, the $i$-th iteration can be executed, and thus the remaining
operations in $C_k$ are as required.

It follows from the inductive hypothesis that the number of available
$-S_{\alpha}$ in $C'_1,\ldots,C'_{n+1}$ is $\ell-m_{i-1}$.  By the
above case analysis we see that the number is exactly $\ell-m_i$ after
executing the $i$-th iteration. The claim is proved.
\end{proof}

If $G_1$ has a valid schedule, then by Lemma~\ref{lemma:np-1}(2) we know
$\height{G_0}\leq\ell$. It then follows from the proof of
Lemma~\ref{lemma:np-2} that $G_1$ has a valid schedule executable by
\fname{Pairwise}.  Therefore, we have the following lemma.

\begin{lemma}
\label{lemma:pairwise-2}
$G_1$ has a valid schedule if and only if $G_1$ has a valid schedule
executable by \fname{Pairwise}.
\end{lemma}

\subsection{Second Step}
\label{subsec:second-step}
In this subsection we show that the $G_1$ constructed in the first step
can be simulated by another chain graph $G_2$, which uses only two
semaphores, $T_1$ and $T_2$. $G_2$ has $2n+3$ chains. The first chain,
denoted $C_0$, is composed of two $-T_1$'s and two $-T_2$'s.  The
remaining $2n+2$ chains are obtained from those of $G_1$ as follows.
We replace every operation $-S_i$ (and $+S_i$) by a {\em unit} $-U_i$
(and $+U_i$) for each $1\leq i\leq n+2$.  Each unit, $-U_i$ or $+U_i$,
is a sequence of operations on $T_1$ and $T_2$, as shown in
Figure~\ref{fig:two-semaphores}\note{Figure~\ref{fig:two-semaphores}}.  
We also denote those $2n+2$ chains of
$G_2$ by $C_1,\ldots,C_{n+1}$ and $C'_1,\ldots,C'_{n+1}$. Clearly, $G_2$
can be constructed in polynomial time. 
\begin{figure}%[p]
  \begin{center}
    \leavevmode
    \begin{displaymath}
      \begin{array}[t]{l}
        +T_1\\+T_2\\
        \left.\!\!\!
        \begin{array}[r]{c}
          -T_1\\+T_2\\
          \vdots\\
          -T_1\\+T_2
        \end{array}
        \right\}\mbox{$i+1$ pairs}\\
        -T_2\\-T_2\\-T_2\\-T_2
      \end{array}\qquad
      \begin{array}[t]{l}
        +T_1\\+T_2\\
        \left.\!\!\!
        \begin{array}[r]{c}
          +T_1\\-T_2\\
          \vdots\\
          +T_1\\-T_2
        \end{array}
        \right\}\mbox{$i+1$ pairs}\\
        +T_2\\+T_2\\-T_1\\-T_1
      \end{array}
    \end{displaymath}
    \caption{The sequence of operations for a $-U_i$ is at left and that
      for a $+U_i$ is at right, for any $1\leq i\leq n+2$.}
    \label{fig:two-semaphores}
  \end{center}
\end{figure}

Note that the sequence of operations in each unit is arranged such that
only a $-U_i$ and a $+U_i$ can ``unlock'' each other. To be more
specific, suppose each of $T_1$ and $T_2$ has initial value -2, which
will be the case if the four operations in $C_0$ are executed.  Consider
a graph $U_{ij}$ for some $1\leq i,j\leq n+2$ composed of two units,
$-U_i$ and $+U_j$, each forms a single chain. One can easily verify that
$U_{ij}$ has a valid schedule if $i=j$. 
%(In fact it also holds for the
%other direction. We do not emphasize it, however, because it is not that
%relevant to our proof.) 
Moreover, after executing all the operations of $U_{ii}$, the values of
$T_1$ and $T_2$ go back to $-2$.

We claim that $G_1$ has a valid schedule if and only if $G_2$ has a
valid schedule. The only-if part is straightforward.  Suppose $G_1$
has a valid schedule.  By Lemma~\ref{lemma:pairwise-2}, $G_1$
has a valid schedule executable by \fname{Pairwise}.  Note that we can
execute the four operations of $C_0$ first, which decrease the value
of both semaphores down to -2. Clearly, the remaining $2n+2$ chains of
units can be completely pairwisely executed by following the sequence
of corresponding operations in $G_1$ executed by \fname{Pairwise}.
Therefore, $G_2$ has a valid schedule.

It takes some added work to prove the other direction of the above
claim.  A unit is {\em active} if its third operation is executed.  A
unit is {\em finished} (and thus inactive) if its fifth-to-last
operation is executed.  Suppose $G_2$ has a valid schedule. Consider
the sequence of the units of $G_2$ that become active in the valid
schedule.  It follows from the following lemma that the corresponding
sequence of operations of $G_1$ is a valid schedule of $G_1$. In fact
it is ``pairwise'', since in the schedule each $-S_i$ is immediately
followed by a $+S_i$.

\begin{lemma}
\label{lemma:two-semaphores}
Consider the execution of a valid subschedule.  
\begin{enumerate}
\item When there is no
active unit, the next unit that becomes active must be a $-U_i$ for
some $1\leq i\leq n+2$.
\item Before that active $-U_i$ is finished, a
$+U_i$ must become active.
\item No unit will become active unless these
two active units are finished.
\end{enumerate}
\end{lemma}

\begin{proof}
At the beginning of the valid schedule, no unit is active. We show the
first statement of the lemma holds.  At this moment there are two
$-T_1$'s and two $-T_2$'s available (in $C_0$). They are our only hope
for activating any unit, since each unit is guarded by two $+T_1$'s
and two $+T_2$'s.  Assume for a contradiction that the first unit
becoming active is a $+U_i$ for some $1\leq i\leq n+2$.  Note that as
soon as the first $+U_i$ becomes active, at least two $+T_1$'s are
already executed. Since at most two $-T_1$'s are executed so far,
there is no way to activate any other unit. The execution thus cannot
proceed.

When the first unit $-U_i$ becomes active, one can see that the second
statement of the lemma holds by verifying the following.
\begin{itemize}
\item The active $-U_i$ will not be finished unless another unit
becomes active, since otherwise the execution will be blocked by
some $+T_2$'s.
\item The next active unit must be a $+U_j$ for some $1\leq j\leq
n+2$, since otherwise the execution will be blocked by some
$+T_2$'s.
\item if $i<j$, the execution will be blocked by some $+T_1$'s. If
$i>j$, then the execution will be blocked by some $+T_2$'s. Therefore, the
next active unit must be a $+U_i$.
\end{itemize}

When those two units are active, in order to activate other units, we
can only hope for the $-T_1$'s at the end of the active $+U_i$. In
order to reach those $-T_1$'s, the preceding consecutive $+T_2$'s must
be penetrated.  Hence, at least two $-T_2$'s at the end of the active
$-U_i$ must be executed first. Therefore, those two active units $-U_i$
and $+U_i$ must be finished before any other unit becomes active.
This confirms the third statement of the lemma.

Note that as soon as the active $+U_i$ is finished (and so must be the
active $-U_i$), the situation is exactly the same as the situation at
the very beginning of the execution. Namely we have two $-T_1$'s and
two $-T_2$'s available, which are again our only hope for activating
any other units.  Therefore, all the above argument follows
inductively.  The lemma is proved.
\end{proof}

\subsection{Third Step}
Let $v$ be the first operation of the $C_0$ in $G_2$. Let $w$ be the
last operation of the $C_{n+1}$ in $G_2$.  We claim that $v\chb w$ if
and only if $G_2$ has a valid schedule.  note that $v$ is always the
first node in any valid subschedule of $G_2$. The if-part of the claim
holds trivially. It remains to prove the only-if-part of the claim.

Let $X$ be a valid subschedule of $G_2$ in which $v$ precedes $w$. 
Consider the sequence of the units of $G_2$ that become active while
executing $X$.  It follows from Lemma~\ref{lemma:two-semaphores} that
the corresponding sequence of operations of $G_1$ is a valid subschedule
of $G_1$, which definitely contains the last $+S_{\alpha}$ of the
$C_{n+1}$ in $G_1$. Therefore, $G_1$ has a valid schedule by
Lemmas~\ref{lemma:np-1}(2) and~\ref{lemma:np-2}.  Finally it follows
from the claim in \S\ref{subsec:second-step} that $G_2$ has a
valid schedule.

\end{document}